\documentclass[pra,aps,floatfix,footinbib,notitlepage,onecolumn,superscriptaddress,longbibliography]{revtex4-2}
\usepackage[makeroom]{cancel}
\usepackage[euler]{textgreek}
\usepackage[bbgreekl]{mathbbol}
\usepackage{graphicx}
\usepackage[explicit]{titlesec}
\usepackage[utf8]{inputenc}
\usepackage{amsmath,amssymb,mathtools}
\usepackage{bbold}
\usepackage{graphicx}
\usepackage{bm,color}
\usepackage{natbib,hyperref}
\usepackage[capitalise]{cleveref}
\usepackage{enumitem}
\usepackage{adjustbox}
\usepackage{mathrsfs}
\usepackage[normalem]{ulem}
\usepackage{xspace}
\usepackage[dvipsnames]{xcolor}
\usepackage[mathlines]{lineno}
\usepackage{subcaption}

\hypersetup{breaklinks=true,colorlinks=true,linkcolor=blue,urlcolor=blue,citecolor=blue}

\newcommand{\ket}[1]{\left| #1 \right\rangle}

\DeclarePairedDelimiterX\braket[2]{\langle}{\rangle}{#1 \delimsize\vert #2}
\newcommand{\abs}[1]{\left| #1 \right|}
\newcommand{\ketbra}[2]{\ensuremath{|#1 \vphantom{#2} \rangle \langle #2 \vphantom{#1} | }}

\def\O{{\cal{O}}}
\def\t{\theta}
\def\X{\bar{X}}
\def\P{\bar{P}}
\def\D{\bar{D}}

\def\x{\bar{x}}
\def\p{\bar{p}}

\def\hh{\hat{h}}
\def\F{{\cal{F}}}

\def\ts{\tilde{\sigma}}

\DeclareMathOperator\erf{erf}
\DeclarePairedDelimiter\floor{\lfloor}{\rfloor}
\DeclarePairedDelimiter\ceil{\lceil}{\rceil}

\begin{document}
\title{Multi-qubit entanglement generation with squeezed  modes}
\author{Alexandru Macridin}
\affiliation{Fermi National Accelerator Laboratory, Batavia, IL, 60510, USA}
\author{Andrew Cameron}
\affiliation{Fermi National Accelerator Laboratory, Batavia, IL, 60510, USA}
\author{Cristian Pena}
\affiliation{Fermi National Accelerator Laboratory, Batavia, IL, 60510, USA}
\author{Si Xie}
\affiliation{Fermi National Accelerator Laboratory, Batavia, IL, 60510, USA}
\author{Raju Valivarthi}
\affiliation{Division of Physics, Mathematics and Astronomy, California Institute of Technology, Pasadena, California 91125, USA}
\author{Panagiotis Spentzouris}
\affiliation{Fermi National Accelerator Laboratory, Batavia, IL, 60510, USA}

\begin{abstract}
We present a hybrid continuous variable-discrete variable entanglement generation protocol using linear optics and homodyne measurements, capable of producing multiple high-fidelity Bell pairs per protocol iteration, with an approximate $0.5$ success probability. The effectiveness of the protocol is determined by the squeezing strength. To increase the number of Bell pairs, approximately 3 dB of extra squeezing is needed for each additional Bell pair. The protocol also generates single Bell pairs with an approximate $0.75$ probability for squeezing strengths $\lessapprox 15dB$, achievable with current technology.
\end{abstract}

\maketitle

\section{Introduction}
\label{sec:intro}

Emerging quantum technologies, such as quantum networks and the quantum internet, require entanglement distribution over long distances and at high rates. Most entanglement distribution methods rely fundamentally on entanglement generation protocols (EGPs) and entanglement swapping protocols (ESPs)~\cite{zukowski_prl_1993, pan_prl_1998, Look_PRA_1999}.
In this paper, we introduce a hybrid continuous-variable (CV)–discrete-variable (DV) entanglement generation protocol that employs squeezed optical modes, linear optics, and homodyne measurements. In this protocol, local entanglement between squeezed light modes and multi-qubit nodes is swapped via homodyne measurement of the optical modes, resulting in the creation of multiple Bell pairs per iteration between spatially separated multi-qubit nodes. The protocol can be readily integrated into subsequent quantum repeater and networking schemes.

The most feasible way for current technology to implement an ESP is by using linear optical elements to transmit and measure information encoded in photonic degrees of freedom. The common example is the implementation of an ESP with photonic qubits~\cite{pan_prl_1998}, which requires a Bell state measurement (BSM). However, one major drawback of the ESP with photonic qubits and linear optics is that the BSM is non-deterministic, leading to a success probability that does not exceed $0.5$~\cite{Lutkenhaus_pra_1999,Vaidman_PRA_1999,Calsamiglia_apb_2001}. Additionally, quantum networks are plagued by signal loss across optical fibers, BSM errors, memory decoherence, classical communication delays, and other issues. All these factors further decrease the success probability and limit the Bell pair transmission rate.

A large research effort is currently underway to increase the entanglement generation rate. This is being pursued by developing new networking protocols using existing entanglement swapping and distillation protocols (for example, multiplexing and parallelization schemes), or by developing new EGP methods. There are many proposals in the literature to increase the success probability or to overcome the non-deterministic nature of an EGP with optical modes, including photon hyperentanglement~\cite{Kwiat_pra_1998}, non-linear photonic interactions~\cite{kim_prl_2001,Barrett_pra_2005,Jian_cpl_2009,Bai_2011,Wang_QIP_2017}, optical squeezing~\cite{zaidi_prl_2013}, or the inclusion of ancillary photonic qubits~\cite{Ewert_prl_2014,Matthias_SA_2023}.
Hybrid DV-CV protocols, which leverage the interaction between qubits and a coherent optical mode combined with homodyne measurement~\cite{vanLoock_PRL_2006}, cat-state projection, and single-photon detection~\cite{Munro_prl_2008}, have been proposed for the nondeterministic generation of a single Bell pair and subsequently used in quantum repeaters. Unlike these hybrid approaches, our hybrid DV-CV protocol utilizes squeezed light instead of coherent light and enables entanglement between multiple pairs of qubits per homodyne measurement and subsequent local operations.

We present a hybrid DV-CV nondeterministic protocol for generating multiple Bell pairs between two distant nodes. At two different locations, a multi-qubit node is entangled with a squeezed CV mode.
The two CV modes are then transmitted to a third location, where they pass through a $50:50$ beam splitter, and their quadratures are measured using homodyne detection. Local qubit operations, dependent on the measurement outcome, are then performed at each node.
 Unlike common photonic-qubit-based protocols, which generate one Bell pair per transmitted and measured signal, our protocol entangles multiple ($n_b$) Bell pairs per protocol iteration with a finite success probability. The effectiveness of the protocol is determined by the squeezing strength. The success probability decreases with decreasing squeezing. For a fixed success probability, the required squeezing, measured in decibels, scales linearly with
$n_b$. We find that  $10\log_{10}\left( 2\right) dB \approx 3.01 dB$ of extra squeezing is needed for each additional Bell pair. At sufficiently large squeezing and for large
$n_b$, the success probability is approximately $0.5$.
The protocol can also be used to create a single Bell pair. In this case, with squeezing values readily achievable with current technology, the success probability is approximately $0.75$, which is significantly higher than that achieved by protocols using photonic qubits and incomplete BSM.
Note that the protocol proposed in Ref~\cite{Ewert_prl_2014}, which employs ancillary photonic qubits, also has a 0.75 theoretical success probability. For practical implementations with current technology, a key difference between the protocol in Ref.~\cite{Ewert_prl_2014} and ours is that they employ photon-number-resolving measurements, whereas we rely on homodyne measurements, which may prove to be more experimentally feasible.

Both proposed applications of our protocol, generating  multiple Bell pairs per protocol iteration with a $0.5$ success probability and generating a single Bell pair with a $0.75$ probability, have the potential to significantly improve entanglement generation rates when implemented in a network. A direct comparison of the achievable improvement with that of previous protocols is of great interest. However, in addition to the EGP, the actual entanglement generation rate is influenced by optical fiber losses, various potential errors, and specific quantum repeaters and network protocols. We do not consider these effects here; the main goal of this paper is to introduce the protocol steps in an ideal scenario without losses. We plan to address loss and other realistic effects in future work.

The paper is organized as follows: \Cref{sec:bkg} provides the necessary theoretical background. The CV and DV variables are introduced in~\cref{sec:cvdv}, followed by a presentation of a pure CV entanglement swapping protocol in~\cref{sec:cv_protocol}. The hybrid protocol is described in~\cref{sec:es_protocol}. In~\cref{sec:purif}, a detailed analytical and numerical investigation of the protocol with squeezed vacuum modes is presented. Potential limitations of the protocol, particularly those related to practical implementation in the near future, comparisons with other protocols, and the possible implementation of our protocol in a quantum repeater are discussed in~\cref{sec:disc}. Finally, the summary and conclusions are outlined in~\cref{sec:conc}.

\section{Preliminaries}
\label{sec:bkg}

\subsection{CV and DV variables}
\label{sec:cvdv}

Our protocol involves multi-qubit registers coupled to optical modes. In this section, we introduce the CV variables used to describe the optical modes and the DV variables used to describe the multi-qubit registers.

\subsubsection{CV modes}
\label{ssec:cvmodes}

Continuous modes are vectors belonging to the Hilbert space of square integrable functions, $L^2(\mathbb{R})$, usually represented in the continuous bases made up of the eigenvectors of
the quadrature operators. The quadrature operators, denoted here by $X$ and $P$,
obey the canonical commutation relations $\left[X,P\right]=i$, and are characterized by continuous spectra
\begin{align}
\label{eq:cvxeigen}
X\ket{x}&=x\ket{x}\\
\label{eq:cvpeigen}
P\ket{p}&=p\ket{p},
\end{align}
\noindent with $x,p \in \mathbb{R}$. The eigenvectors $\{\ket{x}\}$ and $\{\ket{p}\}$  are connected by the  Fourier transform
\begin{align}
\label{eq:ftdef}
\ket{p}&=\frac{1}{\sqrt{2 \pi}}\int_{-\infty}^{\infty}    d x e^{i p x} \ket{x}.
\end{align}

\subsubsection{Multi-qubit registers}
\label{ssec:nregs}

A register of $n$ qubits defines a finite $2^n$-dimensional Hilbert space.
This finite Hilbert space, together with the discrete quadrature operators introduced in this section, constitutes a model for the discretization of continuous modes.

The computational basis of the finite Hilbert space is formed by the vectors $\{\ket{j}\}_j$:
\begin{align}
\label{eq:jbasis}
\ket{j}=\ket{j_0}\ket{j_1}...\ket{j_{n-1}},
\end{align}
\noindent  where $\ket{j_q}$ with $j_q \in \{0,1\}$ is the
state of the qubit with index $q \in \{0,...,n-1\}$. The integer
$j \in \left[0,2^n-1\right]$ is defined as
the decimal representation of the binary number $j_0 j_1 ... j_{n-1}$,
\begin{align}
\label{eq:jbinary}
j=\sum_{q=0}^{n-1} j_q 2^{n-1-q}.
\end{align}

We define the discrete quadrature operators acting on the $n$-qubit register
such:
\begin{align}
\label{eq:Xd}
\X&=-\sum_{q=0}^{n-1}  \t_q \sigma^z_q,~\text{ with }~
\t_q= 2^{n-2-q}\Delta, \\
\label{eq:Pd}
\P&= \F \X \F^{-1},
\end{align}
\noindent where $\sigma^z_q=\ketbra{0}{0}_q-\ketbra{1}{1}_q$ is the Pauli $\sigma^z$ operator
acting on qubit $q$, $\Delta$ is the discretization interval defined as
\begin{align}
\label{eq:Delta}
\Delta=\sqrt{\frac{2 \pi}{2^n}},
\end{align}
\noindent and $\F$ is the quantum Fourier transform (QFT)
\begin{align}
\label{eq:QFT}
\F= \frac{1}{2^\frac{n}{2}} \sum_{j,k=0}^{2^n-1} e^{ i \frac{2\pi}{2^n} \left( j-\frac{2^n-1}{2}\right)\left( k-\frac{2^n-1}{2}\right)} \ketbra{k}{j}.
\end{align}
The operator $\X$ acts on the computational  basis as
\begin{align}
\label{eq:Xxj}
\X \ket{j}&=\x_j \ket{j},~\text{ with }~\x_j=\left(j-\frac{2^n-1}{2}\right) \Delta.
\end{align}
\noindent Thus, the computational basis vectors are eigenvectors of $\X$, with the eigenvalues forming a set of equidistant points $\{x_j\}_j$ separated by the interval $\Delta$.
\Cref{eq:Xxj} is a discretized version of~\cref{eq:cvxeigen}.

Moreover, the set $\{\F\ket{j}\}_j$ constitutes a basis for the discrete quadrature operator $\P$,
\begin{align}
\label{eq:Pxj}
\P \left(\F\ket{j}\right)&= \p_j \left(\F\ket{j}\right),~\text{ with }~\p_j=\left(j-\frac{2^n-1}{2}\right) \Delta,
\end{align}
\noindent with the eigenvalues forming a set of equidistant points separated by the interval $\Delta$. \Cref{eq:Pxj} is the  discretized version of~\cref{eq:cvpeigen}, owing to the fact
that the quantum Fourier transform $\F$ (\cref{eq:QFT}) is the discretized version
of the continuous Fourier transform defined by~\cref{eq:ftdef}.

A detailed  analysis of the discretization of the continuous modes and their representation on a finite Hilbert space is presented in~\cite{macridin_pra_2021,macridin_pra_2024}. However, for the purpose
of this paper, the definition of the discrete quadrature operators given by~\cref{eq:Xd,eq:Pd} will suffice.

In analogy with the CV displacement operator, we  introduce the discrete displacement operator here
\begin{align}
\label{eq:Pdisp}
\D(t)=e^{-i t \Delta \P} \ket{j}&=\ket{\left(j+t\right) \bmod 2^n},
\end{align}
\noindent where $t$ is an integer.

The main goal of the protocol is to produce Bell pairs shared between two separate registers,
for example, one held by Alice and one held by Bob. Considering the Bell state
$\ket\phi^{+}=\left(\ket{00}_{AB}+\ket{11}_{AB}\right)/\sqrt{2}$ and employing~\cref{eq:jbasis},
the state of shared $n$ Bell  pairs is
\begin{align}
\label{eq:nBell}
\ket{\bm{\phi}\left(n\right)}
= \frac{1}{2^\frac{n}{2}} \left(\sum_{j_0=0,1}
\ket{j_0 j_0}_{AB}\right) \otimes \left(\sum_{j_1=0,1}\ket{j_1 j_1}_{AB}\right)\otimes ... \otimes\left(\sum_{j_{n-1}=0,1}\ket{j_{n-1} j_{n-1}}_{AB}\right)
=\frac{1}{2^\frac{n}{2}} \sum_{j=0}^{2^n-1} \ket{jj}_{AB}.
\end{align}

\subsubsection{CV and DV coupling}

\begin{figure}[tb]
    \begin{center}
        \includegraphics*[width=5in]{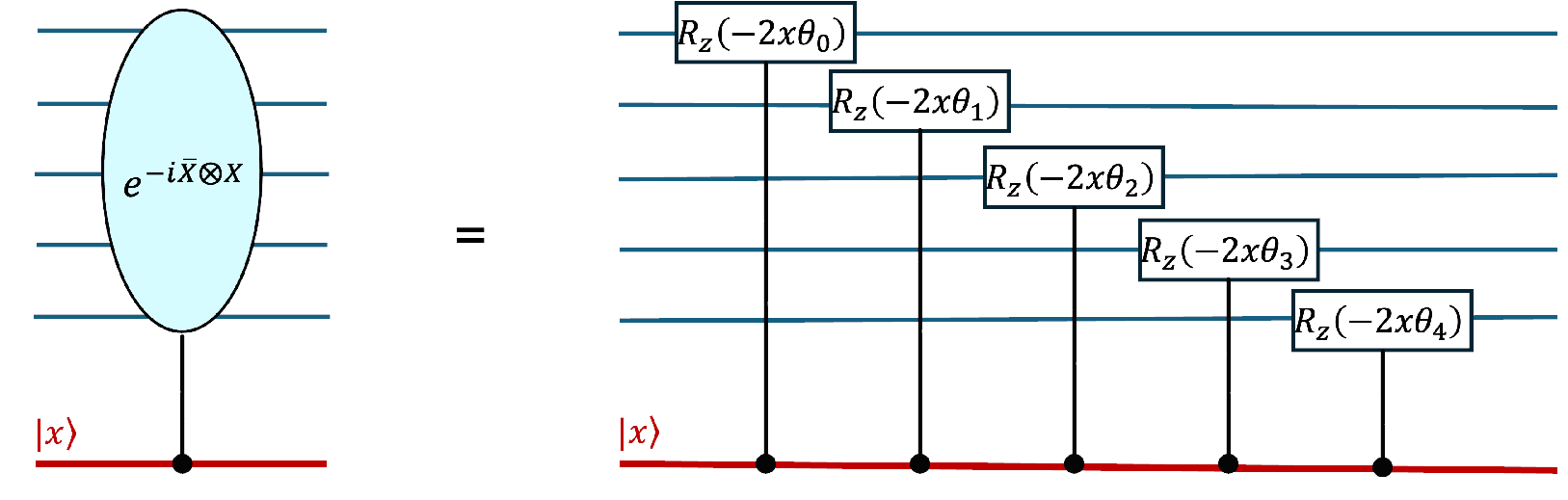}
        \caption{Implementation of the unitary operator $e^{-i \X \otimes X}$ employed by our protocol. The blue lines represent a register of $n$ qubits (in this example, $n=5$), and the red line represents a CV mode expressed in the continuous basis $\{\ket{x}\}$. According to~\cref{eq:cpaseCD}, the implementation requires interaction $H_{int} \propto X \sigma^z_q$ between the CV mode and each qubit. It reduces to $n$ conditional qubit phase rotations with angle $-2 x \t_q$ for $q \in \{0, \dots, n-1\}$, where $\theta_q$ is defined in~\cref{eq:Xd}    and $R_z(\t)=e^{-i \frac{\t}{2} \sigma^z}$ is the single-qubit $z$-rotation.}
        \label{fig:expxx}
    \end{center}
\end{figure}

The hybrid entanglement generation protocol requires a method to
entangle the CV modes with the qubits. As described in the next section,
our protocol employs the unitary
\begin{align}
\label{eq:cpaseCD}
e^{-i \X \otimes X}=\prod_{q=0}^{n-1} e^{i \t_q  \sigma^z_q \otimes  X },
\end{align}
\noindent where $\X$ (see ~\cref{eq:Xd}) acts on $n$-qubit registers and $X$ acts on the CV mode.
The implementation of this unitary is shown in~\cref{fig:expxx} and can be achieved by coupling  the CV mode to each qubit independently, through an interaction proportional to the CV mode quadrature and a qubit Pauli operator.
For example, an interaction Hamiltonian  $H_{int} \propto X \sigma^x_q$, sandwiched between two single-qubit Hadamard gates $H_q$
\begin{align}
e^{i \t_q \sigma^z_q \otimes X  } =H_q e^{i \t_q \sigma^x_q \otimes  X } H_q,
\end{align}
\noindent can be used to implement the unitary~\cref{eq:cpaseCD}.
This kind of interaction is realized, for instance, in systems with transmons coupled to a microwave cavity \cite{Bishop2009}, or in systems with an electromagnetic mode coupled to qubits \cite{Walther2006, Cottet2017, Blais2021}.

\subsection{CV entanglement swapping protocol}
\label{sec:cv_protocol}

Our hybrid protocol is a modified version of the entanglement swapping CV protocol described in~\cite{Look_PRA_1999}. We briefly outline the pure CV protocol below in a form slightly different from the original version, but more suitable for modification into our hybrid version (see also~\cref{app:CVent} for details). Since the purpose of this section is to establish a parallel between the CV protocol and the hybrid one, we will only consider the CV protocol in the nonphysical case that utilizes infinitely squeezed modes.

The CV protocol starts with Alice having two infinitely squeezed modes, $1$ and $2$,
\begin{align}
\ket{\chi}_A=\int \ket{x_1}_A dx_1 \otimes  \int \ket{x_2}_A dx_2.
\end{align}
Next, these two modes are entangled by a gate $e^{-X_1 \otimes X_2}$, where $X_1$ and $X_2$ are the quadrature operators for modes $1$ and $2$,  respectively. At a separate location, Bob entangles modes $3$ and $4$ in the same way. Then Alice sends mode $2$ and Bob sends mode $3$ to Charlie. Charlie sends the two received modes through a $50:50$ beam splitter and, afterwards, measures the quadrature $X$ of one mode and the quadrature $P$ of the other mode.
The outcome of the measurement is sent through classical channels to Alice and Bob.
Alice and Bob employ local operations that are dependent on the measurement results received from Charlie and obtain the entangled state
\begin{align}
\label{eq:cvfinstate}
\ket{\chi}_{AB}=\int \ket{x,x}_{AB} dx.
\end{align}

Our hybrid protocol is inspired by the observation that
the discretized version of~\cref{eq:cvfinstate} corresponds to an $n$-pair Bell state described by~\cref{eq:nBell}.
Recall that the discretization of CV modes on $n$-qubit registers introduced in~\cref{ssec:nregs} maps the CV vector $\ket{x}$   to the $n$-qubit vector $\ket{j}$.
Therefore, if we substitute the CV modes $1$ and $4$ with $n$-qubit states
we expect the resulting state, in the limit of large $n$, to
have a large overlap with an $n$-pair Bell state.

\section{Hybrid CV-DV entanglement generation protocol}
\label{sec:es_protocol}

\begin{figure}[tb]
    \begin{center}
        \includegraphics*[width=5in]{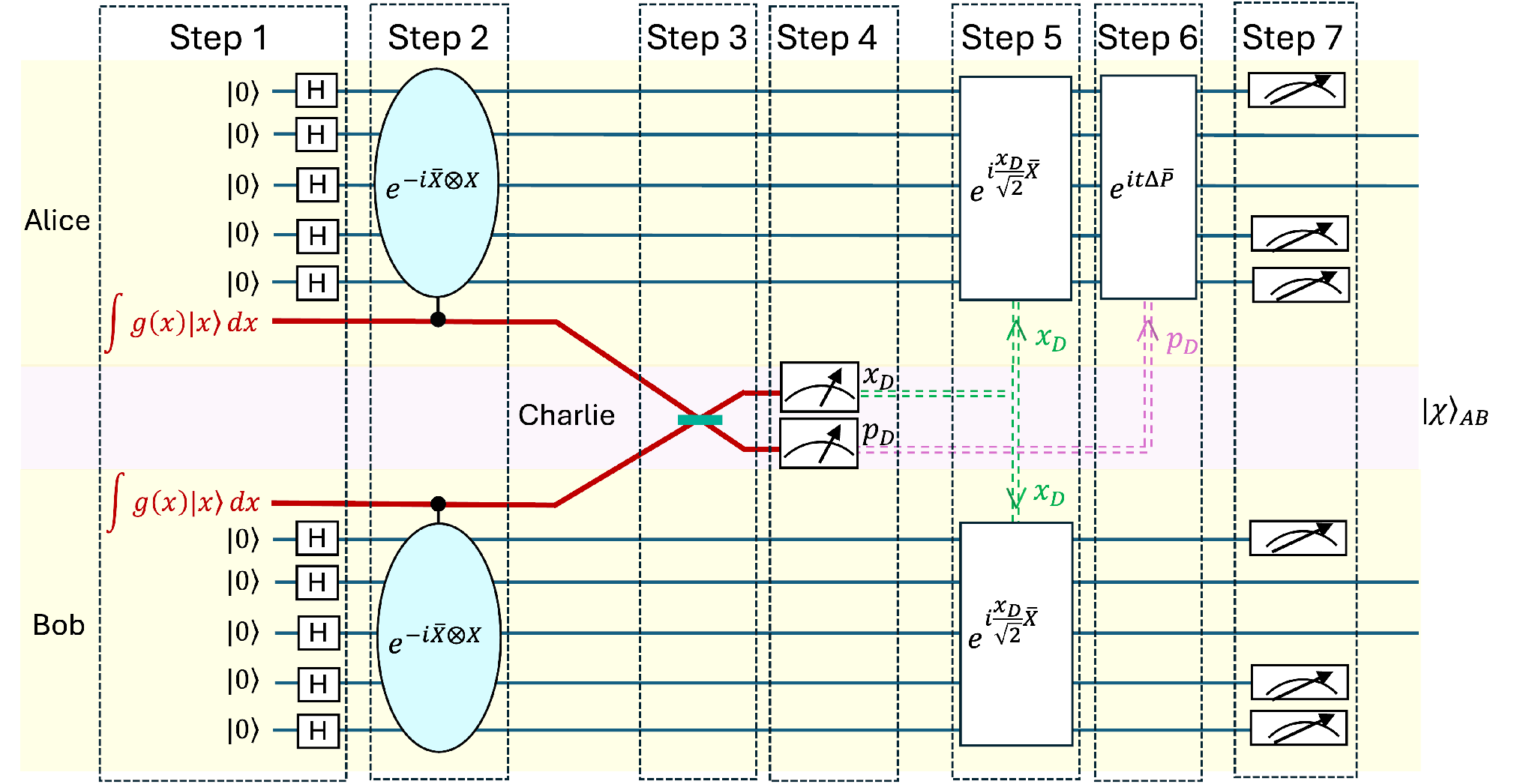}
        \caption{Diagram of the hybrid CV-DV entanglement generation protocol. \emph{Step 1}: Alice and Bob each start with an $n$-qubit register (blue solid lines, $n=5$ in this example) and a CV mode (red thick solid lines). A Hadamard gate $H$ is applied to each qubit. The initial CV states are prepared in the quadrature basis as $\int g(x)\ket{x} dx$ (for squeezed vacuum, see~\cref{eq:gfun}). \emph{Step 2}: The $n$-qubit registers and the CV states are entangled locally through the operator $e^{-i \X \otimes X}$, defined by~\cref{eq:cpaseCD} and shown explicitly in~\cref{fig:expxx}. \emph{Step 3}: The continuous modes are sent to Charlie, who passes them through a $50:50$ beam splitter. \emph{Step 4}: Charlie measures the quadrature $X$ of one mode and the quadrature $P$ of the other and relays the measurement outcomes (denoted by $x_D$ and $p_D$) to Alice and Bob via classical channels (dotted double lines). \emph{Step 5}: Alice and Bob perform the local operations $e^{i \frac{x_D}{\sqrt{2}}\X}$, shown explicitly in~\cref{fig:expxdx}, on their qubit registers. \emph{Step 6}: Alice performs the local operation $e^{i t(p_D) \Delta \P}$, shown explicitly in~\cref{fig:exptP} (see~\cref{eq:tdeltapdef} for the definition of $t(p_D)$). \emph{Step 7}: Alice and Bob measure their first qubit and their last $s_c$ qubits (the choice of $s_c$ is described in the text; in this example, $s_c=2$). The protocol is non-deterministic and succeeds when the final state $\ket{\chi}_{AB}$ approximates $n-1-s_c$ Bell pairs shared between Alice and Bob with the desired fidelity.}
        \label{fig:proto}
    \end{center}
\end{figure}

\begin{figure}[htbp]
    \centering
    \begin{subfigure}[t]{0.45\textwidth}
        \centering
        \includegraphics[width=\textwidth,height=3cm]{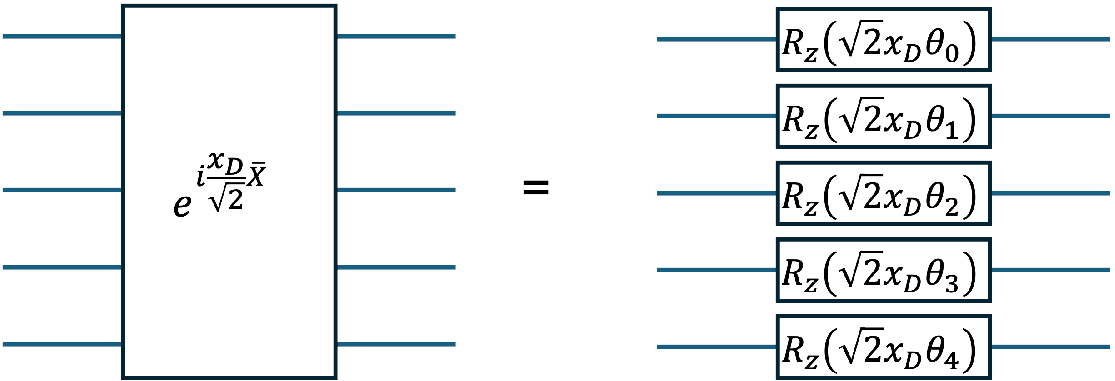}
        \caption{}
        \label{fig:expxdx}
    \end{subfigure}
    \hfill
    \begin{subfigure}[t]{0.45\textwidth}
        \centering
        \includegraphics[width=\textwidth,height=3cm]{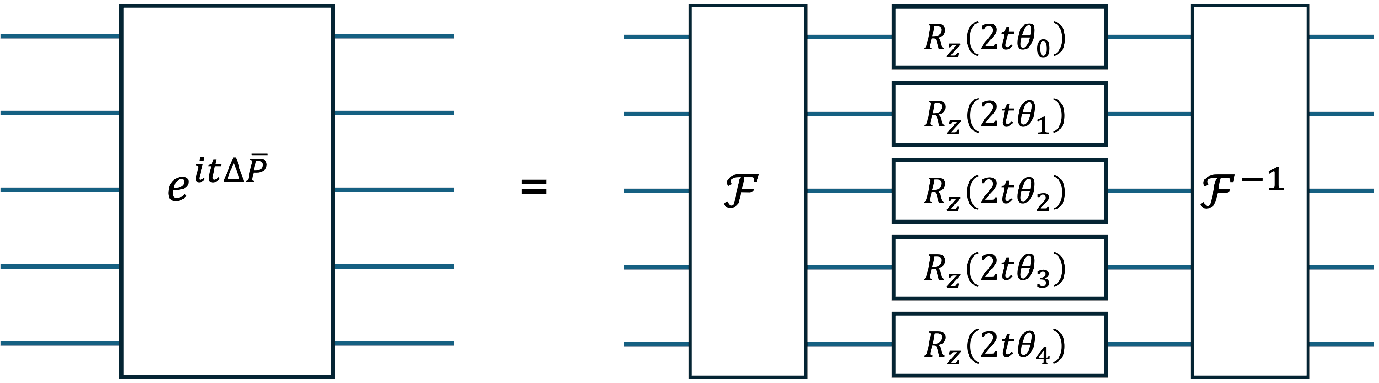}
        \caption{}
        \label{fig:exptP}
    \end{subfigure}
    \caption{a) The implementation of the operation $e^{i \frac{x_D}{\sqrt{2}} \X}$ on a register of $n$ qubits reduces to $n$ single-qubit $z$-rotations.  b) The implementation of the operation $D(-t)=e^{i t \Delta \P}$ on a register of $n$ qubits  reduces to $n$ single-qubit $z$-rotations, sandwiched between a quantum Fourier transform ${\cal{F}}$ and its inverse ${\cal{F}}^{-1}$.  The value of $\t_q$, for $q \in \{0,1,\dots,n-1\}$ is given by~\cref{eq:Xd}. In this example $n=5$. The single-qubit $z$-rotation is defined as $R_z(\t)=e^{-i \frac{\t}{2} \sigma^z}$. }
    \label{fig:mainfig}
\end{figure}

Our protocol describes a method for generating multiple Bell pairs between two distant nodes, each containing  $n$  qubits. The process begins by entangling the qubits at each node with a CV mode. The two CV modes are then transmitted to a third location, where they pass through a $50:50$ beam splitter, and their quadratures are measured using homodyne detection. The outcomes of the homodyne measurements determine the local operations to be subsequently performed at each node. Fidelity can be improved by measuring a subset of qubits in the computational basis. The protocol is non-deterministic, as its success depends on the outcomes of the homodyne and qubit measurements. In this section, a detailed explanation of the protocol steps is provided, along with an analysis of its success probability and fidelity.

The first part of the hybrid protocol is a straightforward modification of the CV protocol
outlined in~\cref{sec:cv_protocol} and~\cref{app:CVent}. In the hybrid protocol, the continuous mode $1$ at Alice and the continuous mode $4$ at Bob are replaced by $n$-qubit registers. The $n$-qubit registers are treated as ``discretized'' continuous modes, as
described in~\cref{ssec:nregs}. The protocol is illustrated in~\cref{fig:proto}, and
consists of the following steps:

\begin{enumerate}
\item
Alice starts with an unentangled state consisting of a direct product
between an $n$-qubit DV state and a CV mode:
\begin{align}
\label{eq:chiiniA}
\ket{\chi\left(\text{step1}\right)}_{A}=\left(\frac{1}{\sqrt{2^n}} \sum_{j=0}^{2^n-1} \ket{j}_A \right)\otimes \left(\int dx_2 g(x_2)\ket{ x_2}_A \right).
\end{align}
\noindent The DV initial state can be obtained by applying a single-qubit Hadamard gate
to every qubit initially prepared in the state $\ket{0}$.
Similarly Bob starts with a product of a CV mode and a $n$-qubit DV state:
\begin{align}
\label{eq:chiiniB}
%\ket{\chi}^1_{B}=\frac{1}{\sqrt{2^n}} \sum_{k=0}^{2^n-1} \int dx_3 g(x_3)\ket{x_3, k}_B.
\ket{\chi\left(\text{step1}\right)}_{B}=  \left(\int dx_3 g(x_3)\ket{ x_3}_B\right) \otimes \left(\frac{1}{\sqrt{2^n}} \sum_{k=0}^{2^n-1} \ket{k}_B\right).
\end{align}
The initial CV modes are defined by the function $g(x)$, with the normalization condition
$\int \abs{g(x)}^2 dx=1$.

\item
Alice and Bob entangle their CV and DV modes by applying the local operator $e^{-i \X \otimes X}$
(see~\cref{eq:Xxj,eq:cpaseCD}):
\begin{align}
\ket{\chi\left(\text{step2}\right)}_{A}=\frac{1}{\sqrt{2^n}} \sum_{j=0}^{2^n-1} \int dx_2 g(x_2) e^{-i \x_j x_2}\ket{j, x_2}_A,
\end{align}
\begin{align}
\ket{\chi\left(\text{step2}\right)}_{B}=\frac{1}{\sqrt{2^n}} \sum_{k=0}^{2^n-1} \int dx_3 g(x_3)e^{-i \x_k x_3}\ket{x_3, k}_B.
\end{align}
\item
Alice and Bob send their continuous modes to Charlie. Charlie sends them through a $50:50$ beam splitter,
\begin{align}
\label{eq:st3}
\ket{\chi\left(\text{step3}\right)}_{ABC}=\frac{1}{2^n} \sum_{j,k=0}^{2^n-1}\int  dx_2 dx_3 g(x_2) g(x_3) e^{-i \x_j x_2} e^{-i x_3 \x_k} \ket{j, k}_{AB}  \ket{\frac{x_2+x_3}{\sqrt{2}}, \frac{-x_2+x_3}{\sqrt{2}}}_C.
\end{align}
\item
Employing homodyne measurement, Charlie measures the quadrature $X$ for one mode and the quadrature $P$ for the other mode. By denoting the results of the  measurement by $x_D$
and $p_D$, the state becomes (see~\cref{app:homodyne}):
\begin{align}
\label{eq:st4}
\ket{\chi\left(\text{step4}\right)}_{AB}&=\frac{1}{2^n \sqrt{P\left(x_D,p_D\right)}}
\sum_{j,k=0}^{2^n-1}
e^{-i x_D \frac{\x_j+\x_k }{\sqrt{2}}}  \hh_{x_D}\left(p_D-\frac{\x_j-\x_k}{\sqrt{2}}\right)\ket{j, k}_{AB},
\end{align}
\noindent where
\begin{align}
\label{eq:hhfun}
\hh_{x_D}(p)=\frac{1}{\sqrt{2\pi}}\int dx g\left(\frac{x_D+x}{\sqrt{2}}\right) g\left(\frac{x_D-x}{\sqrt{2}}\right) e^{i  x p},
\end{align}
\noindent and the probability to measure $x_D$ and $p_D$ is
\begin{align}
\label{eq:probxp}
P\left(x_D,p_D\right)=\frac{1}{2^{2n}} \sum_{j,k=0}^{2^n-1} \abs{\hh_{x_D}\left(p_D-\frac{\x_j-\x_k}{\sqrt{2}}\right)}^2.
\end{align}
Charlie sends the values of $x_D$ and $p_D$ to Alice and Bob through classical channels.
\item
 Alice and Bob apply the operator $e^{i \frac{x_D}{\sqrt{2}} \X}$ to their discrete mode,
\begin{align}
\label{eq:st5}
\ket{\chi\left(\text{step5}\right)}_{AB}&=\frac{1}{2^n \sqrt{P\left(x_D,p_D\right)}}
\sum_{j,k=0}^{2^n-1}
  \hh_{x_D}\left(p_D-\frac{\x_j-\x_k}{\sqrt{2}}\right)\ket{j, k}_{AB}.
\end{align}
Since the operator $\X$ is defined as a linear combination of Pauli $\sigma_q^z$ operators (see~\cref{eq:Xd}), the implementation of $e^{i \frac{x_D}{\sqrt{2}} \X}$ operation on a DV register reduces to $n$ single-qubit $R_z$ rotations, one on each qubit, as illustrated in~\cref{fig:expxdx} (see  Appendix D in~\cite{macridin_pra_2024} for the explicit implementation).
\item
Alice applies the discrete displacement operator
$\D(-t)$,
\begin{align}
\label{eq:step6}
\ket{\chi}_{AB} = \ket{\chi\left(\text{step6}\right)}_{AB}&=\frac{1}{2^n \sqrt{P\left(x_D,p_D\right)}}
\sum_{j,k=0}^{2^n-1}
  \hh_{x_D}\left(p_D-\frac{\x_j-\x_k}{\sqrt{2}}\right)\ket{ \left(j-t\right)\bmod2^n, k }_{AB},
\end{align}
\noindent where the integer parameter $t$
is defined such that
\begin{align}
\label{eq:tdeltapdef}
\sqrt{2} p_D=(t+\delta_p) \Delta,~\text{ with  }~ -0.5<\delta_p \le 0.5 ~\text{ and }~ t \in \mathbb{Z}.
\end{align}
Both the integer $t \left(p_D\right)$ and the real $\delta_p\left(p_D\right)$ are uniquely determined by the measurement outcome $p_D$.
The implementation of $D(-t)=e^{i t  \Delta \P}=\F e^{i t  \Delta \X} \F^{-1}$ (see~\cref{eq:Pdisp,eq:Pd})
requires single-qubit $R_z$ rotations sandwiched between
two quantum Fourier transform (QFT) operations, as illustrated in~\cref{fig:exptP} (see  Appendix C in~\cite{macridin_pra_2024}
for the explicit implementation).

\item
This final step involves measurement of qubits in the computational basis, and can be considered a purification process. At this step Alice and Bob measure the first qubit and the last $s_c$ qubits of their registers. The resulting state is a $2 \times n_b$ qubit state, with $n_b=n-1-s_c$.
The fidelity of the protocol is
\begin{align}
F_{n_b}=\abs{\braket{\chi}{\bm{\phi}(n_b)}}^2,
\end{align}
\noindent where $\ket{\bm{\phi}(n_b)}$ is the $n_b$-pair Bell state defined by~\cref{eq:nBell}
on the qubits $\{1,\dots,n-1-s_c\}$ of both Alice and Bob.
The number of qubits necessary to be measured (equal to $1+s_c$)
depends on the wavefunction of the initial CV modes (for example, on the squeezing strength when the wavefunction is Gaussian), the size $n$ of the qubit registers, and the desired protocol fidelity. The success of the protocol is probabilistic and depends on the value of the homodyne measurements ($x_D$ and $p_D$) and on the outcome of the measured qubits.

This last step of our protocol will be described at length in~\cref{sec:purif} for the case when
the initial wavefunction $g(x)$ is Gaussian.

\end{enumerate}

\subsection{Entanglement generation  with squeezed vacuum}
\label{sec:purif}

The details of the last step (step $7$) of the hybrid protocol depend on the CV wavefunction.
Here we will present the case with initial Gaussian CV modes,
\begin{align}
\label{eq:gfun}
g(x)&=\pi^{-\frac{1}{4}}\frac{1}{\sqrt{\sigma}}e^{-\frac{x^2}{2 \sigma^2}}.
\end{align}
\noindent The Gaussian wavefunctions describe squeezed vacuum states, with the standard deviation $\sigma$ related to the
squeezing factor through the relation $\sigma=e^{-r}$~\cite{gerry_knight_2004}. In our units, $\sigma=1$ corresponds to the vacuum state. Our protocol requires $\sigma > 1$ ({\emph{i.e.}} $r<0$) which corresponds to the squeezing of the $P$ quadrature.

Employing~\cref{eq:gfun,eq:hhfun} we obtain,
\begin{align}
\hh_{x_D}(p)=\frac{1}{\sqrt{\pi}}e^{-\frac{x_D^2}{2\sigma^2}}e^{-\frac{p^2 \sigma^2}{2}}.
\end{align}

\Cref{eq:probxp} yields the
probability to measure $x_D$ and $p_D$,
\begin{align}
P\left(x_D,p_D\right)=P\left(x_D\right)P\left(p_D\right)
\end{align}
\noindent where
\begin{align}
\label{eq:pxd}
P\left(x_D\right)=\frac{1}{\sigma\sqrt{\pi}}e^{-\frac{x_D^2}{\sigma^2}},
\end{align}
\noindent and
\begin{align}
\label{eq:ppd}
P\left(p_D\right)=
%\frac{\sigma}{\sqrt{\pi}}\frac{1}{2^{2n}} \sum_{j,k=0}^{2^n-1} e^{- \sigma^2\left(p_D-\frac{\x_j-\x_k}{\sqrt{2}}\right)^2}=
\frac{1}{2^{2n}}\sum_{j,k=0}^{2^n-1} \abs{u\left(p_D-\frac{\x_j-\x_k}{\sqrt{2}}\right)}^2,
\end{align}
\noindent with
\begin{align}
\label{eq:up}
u\left(p\right)=\sqrt{\frac{\sigma}{\sqrt{\pi}}}e^{- \frac{\sigma^2}{2}p^2}.
\end{align}
\noindent The measurement probabilities of the two quadratures, $x_D$ and $p_D$,
are independent of each other for Gaussian modes. This is not  necessarily true
for other choices of $g\left(x\right)$.

For Gaussian modes the state given by~\cref{eq:step6} is independent of the outcome $x_D$, and can be written as
\begin{align}
\label{eq:chi_pm}
\ket{\chi}_{AB}
% &=\frac{1}{2^n \sqrt{P\left(x_D,p_D\right)}}
% \sum_{j,k=0}^{2^n-1}
%   \hh_{x_D}\left(p_D-\frac{\x_j-\x_k}{\sqrt{2}}\right)\ket{\x_{\left(j-t\right)~mod~2^n}, \x_k }_{AB}\\ \nonumber
  % &=\frac{1}{\sqrt{P(p_D)}}\sqrt{\frac{\sigma}{\sqrt{\pi}}}\frac{1}{2^{n}}\sum_{j,k=0}^{2^n-1} e^{- \frac{\sigma^2}{2}\left(p_D-\frac{\x_j-\x_k}{\sqrt{2}}\right)^2} \ket{\x_{\left(j-t\right)\bmod{2^n}}, \x_k }_{AB}\\ \nonumber
  &=\frac{1}{\sqrt{P(p_D)}}\frac{1}{2^{n}}\sum_{j,k=0}^{2^n-1} u\left(p_D-\frac{\x_j-\x_k}{\sqrt{2}}\right) \ket{\left(j-t\right)\bmod{2^n}, k }_{AB}.
\end{align}

The purification step of the protocol can be understood by analyzing the following approximation. For a given $\sigma$, there is a positive dimensionless cutoff parameter c such that $u(p) \approx 0$ when $\abs{p} \sigma > c$. In the following analytical expressions, we will neglect $u(p)$ for $\abs{p} > \frac{c}{\sigma}$. We keep $c$ as a parameter to control this truncation error. Increasing the value of $c$ decreases the truncation errors exponentially. As discussed in~\cref{ssec:fidprob}, the truncation errors are related to the protocol fidelity. Numerically, we find that $c \gtrapprox 2.2$ yields fidelity errors $\lessapprox 1\%$ for our protocol.

The truncation of  $u(p)$ implies that in~\cref{eq:chi_pm} we can write
\begin{align}
\label{eq:ucutoff}
u\left(p_D-\frac{\x_j-\x_k}{\sqrt{2}}\right)& \equiv u\left[\left(t-j+k+\delta_p\right)\frac{\Delta}{\sqrt{2}}\right]
\approx
% u\left(\delta_p\frac{\Delta}{\sqrt{2}}\right)\delta_{j-t,k}+u\left[\left(1+\delta_p\right)\frac{\Delta}{\sqrt{2}}\right]\delta_{j-t,k-1}\\ \nonumber
% &+u\left[\left(-1+\delta_p\right)\frac{\Delta}{\sqrt{2}}\right]\delta_{j-t,k+1}
% +u\left[\left(2+\delta_p\right)\frac{\Delta}{\sqrt{2}}\right]\delta_{j-t,k-2}
% +u\left[\left(-2+\delta_p\right)\frac{\Delta}{\sqrt{2}}\right]\delta_{j-t,k+2}\\ \nonumber
% &...+u\left[\left(m+\delta_p\right)\frac{\Delta}{\sqrt{2}}\right]\delta_{j-t,k-m}
% +u\left[\left(-m+\delta_p\right)\frac{\Delta}{\sqrt{2}}\right]\delta_{j-t,k+m}\\ \nonumber
\sum_{m=-m_c}^{m_c} u\left[\left(m+\delta_p\right)\frac{\Delta}{\sqrt{2}}\right]\delta_{j-t,k-m},
\end{align}
\noindent where $t$ and $\delta_p$ are defined by~\cref{eq:tdeltapdef}, $\x_j$ and $\x_k$
by~\cref{eq:Xxj},
and $m_c$ can be any positive integer such that
\begin{align}
\label{eq:sigmamcut}
\left(m_c+\frac{1}{2}\right)\frac{\Delta}{\sqrt{2}}\sigma >c.
\end{align}

It is convenient to replace $m_c$ with $2^{s_c}-1$ in~\cref{eq:ucutoff}, where
the parameter $s_c$ is a positive integer such that $2^{s_c}-1 \ge m_c$.
\Cref{eq:sigmamcut} implies that $s_c$ should be chosen such that
\begin{align}
\label{eq:sigmascut}
\left(2^{s_c}-\frac{1}{2}\right)\sqrt{\frac{\pi}{2^{n}}}\sigma >c.
\end{align}

Finally, employing the truncation described by~\cref{eq:ucutoff}, the wavefunction described by~\cref{eq:chi_pm}  becomes
\begin{align}
\label{eq:chiab}
\ket{\chi}_{AB}
  &\approx
  \frac{1}{2^n\sqrt{P(p_D)}}\sum_{m=-2^{s_c}+1}^{2^{s_c}-1} u\left[\left(m+\delta_p\right)\frac{\Delta}{\sqrt{2}}\right]  \sum_{k=\max(0,m-t)}^{2^n-1-\min(0,m-t)}  \ket{\left(k-m\right)\bmod{2^n}, k }_{AB}.
  \end{align}

\subsubsection{Purification step}
\label{ssec:pstep}

The purification step of our protocol requires the measurement of the first qubit (indexed
by $0$ in our notation, see~\cref{eq:jbasis}) and  the last $s_c$ qubits (indexed by $\{n-s_c, \dots, n-1\}$) at each register. In total each of Alice and Bob measures and discards $1+s_c$ qubits.
Consequently, the total number of Bell pairs for successful protocols will be
\begin{align}
\label{eq:nbdef}
n_b=n-1-s_c.
\end{align}
The optimal value of $s_c$ for our protocol, which depends on the fidelity, the squeezing $\sigma$ and the registers' size $n$, will be discussed in~\cref{ssec:fidprob}.

We begin by analyzing  the measurement of the last qubit in Alice's and Bob's registers, indexed as $n-1$ in our notation. If the measured values of this qubit at Alice and Bob are different, the protocol is considered unsuccessful. If the measured values are the same, {\emph{i.e.}}, either both $0$ or both $1$, the post-measurement state is obtained by applying the projector operators:
\begin{align}
\label{eq:P00}
P_{0_{n-1}0_{n-1}}= \mathbb{I}_{2 \times \left(n-1\right)}\otimes\ketbra{0_{n-1}0_{n-1}}{0_{n-1}0_{n-1}}
\end{align}
\noindent or
\begin{align}
\label{eq:P11}
P_{1_{n-1}1_{n-1}}= \mathbb{I}_{2 \times \left(n-1\right)}\otimes\ketbra{1_{n-1}1_{n-1}}{1_{n-1}1_{n-1}},
\end{align}
\noindent respectively.

In order to calculate how these operators act on the state described by~\cref{eq:chiab},
we used the fact
that an $n$-qubit integer $k$ (see~\cref{eq:jbinary}) can be written as
\begin{align}
\label{eq:kprime}
k=\sum_{q=0}^{n-2} k_q 2^{n-1-q}+k_{n-1}=2\sum_{q=0}^{n^\prime-1} k_q 2^{n^\prime-1-q}+k_{n-1}
=2k^\prime+k_{n-1},
\end{align}
\noindent where  $k^\prime$ is the decimal representation of the state encoded in the first $n^\prime=n-1$ qubits. Then, for even $m=2m^\prime$ and even $k=2k^\prime$ we have,
\begin{align}
\label{eq:Plast00}
P_{0_{n-1}0_{n-1}}\ket{\left(2k^\prime-2m^\prime\right)\bmod{2^n}, 2k^\prime }_{AB}
&=\ket{\left(k^\prime-m^\prime\right)\bmod{2^{n^\prime}}, k^\prime }_{AB}\otimes \ket{0_{n-1}0_{n-1}}_{AB},\\
\label{eq:Plast00p}
P_{1_{n-1}1_{n-1}}\ket{\left(2k^\prime-2m^\prime\right)\bmod{2^n}, 2k^\prime }_{AB}
&=0,
\end{align}
\noindent for even $m=2m^\prime$ and odd $k=2k^\prime+1$,
\begin{align}
\label{eq:Plast11p}
P_{0_{n-1}0_{n-1}}\ket{\left(2k^\prime+1-2m^\prime\right)\bmod{2^n}, 2k^\prime+1 }_{AB}&=0,\\
\label{eq:Plast11}
P_{1_{n-1}1_{n-1}}\ket{\left(2k^\prime+1-2m^\prime\right)\bmod{2^n}, 2k^\prime+1 }_{AB}&=
\ket{\left(k^\prime-m^\prime\right)\bmod{2^{n^\prime}}, k^\prime }_{AB}\otimes \ket{1_{n-1}1_{n-1}}_{AB},
\end{align}
\noindent while for odd $m$ and any $k$,
\begin{align}
\label{eq:Plastmodd}
P_{0_{n-1}0_{n-1}}\ket{\left(k-m\right)\bmod{2^n}, k }_{AB}=
P_{1_{n-1}1_{n-1}}\ket{\left(k-m\right)\bmod{2^n}, k }_{AB}=0.
\end{align}

Employing~\cref{eq:Plast00,eq:Plast00p,eq:Plast11,eq:Plast11p,eq:Plastmodd},
 the unnormalized wavefunction described by~\cref{eq:chiab} becomes, after the measurement,
\begin{align}
\label{eq:prj00}
P_{0_{n-1}0_{n-1}}\ket{\chi}_{AB}
&= \frac{1}{\sqrt{P(p_D)}}\frac{1}{2^n}\sum_{m^\prime=-2^{s_c-1}+1}^{2^{s_c-1}-1}
 u\left[\left(2m^\prime+\delta_p\right)\frac{\Delta}{\sqrt{2}}\right] \times \\ \nonumber &\times \sum_{k^\prime=\max(0,m^\prime-\floor{\frac{t}{2}})}^{2^{n-1}-1+\min(0,m^\prime-\floor{\frac{t}{2}})}  \ket{\left(k^\prime-m^\prime\right)\bmod{2^{n-1}}, k^\prime }_{AB} \otimes \ket{0_{n-1}0_{n-1}}_{AB},
\end{align}
\noindent when the value of measured qubits at both Alice and Bob is $0$, or
\begin{align}
\label{eq:prj11}
P_{1_{n-1}1_{n-1}}\ket{\chi}_{AB}
&= \frac{1}{\sqrt{P(p_D)}}\frac{1}{2^n}\sum_{m^\prime=-2^{s_c-1}+1}^{2^{s_c-1}-1}
 u\left[\left(2m^\prime+\delta_p\right)\frac{\Delta}{\sqrt{2}}\right] \times \\ \nonumber &\times \sum_{k^\prime=\max(0,m^\prime-\ceil{\frac{t}{2}})}^{2^{n-1}-1+\min(0,m^\prime-\ceil{\frac{t}{2}})}  \ket{\left(k^\prime-m^\prime\right)\bmod{2^{n-1}}, k^\prime }_{AB} \otimes \ket{1_{n^\prime}1_{n-1}}_{AB},
\end{align}
\noindent  when the value of measured qubits is $1$.
In~\cref{eq:prj00,eq:prj11},  $\ket{\left(k^\prime-m^\prime\right)\bmod{2^{n^\prime}}, k^\prime}_{AB}$ is a state stored
on the first $n-1$ qubits of Alice and the first $n-1$ qubits of Bob.
The limits of the summation over the $k^{\prime}$ in~\cref{eq:prj00} involves the floor integer $\floor{\frac{t}{2}}$, which rounds down $\frac{t}{2}$ (see~\cref{eq:floorf}), while in~\cref{eq:prj11} it involves the ceiling integer $\ceil{\frac{t}{2}}$, which rounds up $\frac{t}{2}$ (see~\cref{eq:ceilf}).
%These two integer truncation operations are different when $t$ is odd.

\Cref{eq:prj00,eq:prj11} can be generalized to describe the
state after the measurement of all $s_c$  qubits indexed by $n-s_c, \dots, n-1$,
in Alice's and Bob's registers. The protocol is considered unsuccessful when the outcome at Alice is not identical to the outcome at Bob. When the outcomes are identical, they determine a  sequence of $s_c$ values  of $0$ and $1$,
\begin{align}
\label{eq:Sdef}
S=\{j_{n-s_c},\dots, j_{n-2},j_{n-1}\},
\end{align}
where $j_{q}$ is the outcome of the measurement of the qubit indexed by $q$
at both Alice's and Bob's locations.
The unnormalized post-measurement state is given by
\begin{align}
\label{eq:prjsctimes}
P_S\ket{\chi}_{AB}
&= \frac{1}{2^n\sqrt{P(p_D)}}
 u\left(\delta_p\frac{\Delta}{\sqrt{2}}\right) \times \sum_{k=\max(0,-t_S)}^{2^{n-s_c}-1+\min(0,-t_S)}  \ket{k, k }_{AB} \otimes \ket{S}_{AB},
\end{align}
\noindent where
\begin{align}
\ket{S}_{AB}&= \ket{j_{n-s_c}j_{n-s_c}}_{AB} \otimes \dots \otimes \ket{j_{n-1}j_{n-1}}_{AB},
\end{align}
\begin{align}
P_S&=\mathbb{I}_{2\times\left(n-s_c\right)}\otimes\ketbra{S}{S},
\end{align}
\noindent and
$\ket{k, k }_{AB}$ is a state spanned by the first $n-s_c$ qubits at Alice and by the first $n-s_c$ qubits at Bob. The integer $t_S$ is defined by
\begin{align}
\label{eq:tS}
t_S= \left[...\left[\left[ t \times \frac{1}{2}\right]_{j_{n-1}} \times \frac{1}{2} \right]_{j_{n-2}} ... \times \frac{1}{2}\right]_{j_{n-s_c}},
\end{align}
\noindent where we denote
\begin{align}
\label{eq:floorf}
\left[x \right]_0 &\equiv \floor{x}\equiv~\text{ floor function}(x)\equiv\max\{n \in \mathbb{Z} | n \le x\}\\
\label{eq:ceilf}
\left[x\right]_1 &\equiv \ceil{x}\equiv~\text{ ceiling function}(x)\equiv\min\{n \in \mathbb{Z} | n \ge x\}.
\end{align}

Now we will analyze the measurement of the first qubit, indexed by $0$, in Alice's and
Bob's registers. We start by imposing the following requirement:
the protocol is considered  successful only when
\begin{align}
\label{eq:pdrange}
p_D \in \left(-\left(2^n+1\right)\frac{\Delta}{2\sqrt{2}}, \left(2^n+1\right)\frac{\Delta}{2\sqrt{2}} \right] \iff  t \in \left[-2^{n-1}, 2^{n-1} \right].
\end{align}
\noindent In the event that the homodyne measurement performed by Charlie yields a value of $p_D$ outside of this range, the protocol is abandoned. \Cref{eq:pdrange} implies
$\abs{t_S}\le 2^{n-s_c-1}$, where $t_S$ was defined by~\cref{eq:tS}. Therefore, for a successful protocol, the state described by~\cref{eq:prjsctimes} can be written as
\begin{align}
\label{eq:chifinal0}
\ket{\chi}_{AB}
&= \frac{1}{2^n\sqrt{P(p_D) P(S)}}
 u\left(\delta_p\frac{\Delta}{\sqrt{2}}\right) \left[\sum_{k=\max(0,-t_S)}^{2^{n-s_c-1}-1}  \ket{k, k }_{AB} + \sum_{k=2^{n-s_c-1}}^{2^{n-s_c}-1+\min(0,-t_S)}  \ket{k, k }_{AB} \right] \otimes \ket{S}_{AB},
\end{align}
\noindent where $P(S)$ is a normalization constant equal to the probability
of the measurement of the $2 \times s_c$ qubits yielding the sequence $S$ at both Alice and Bob locations. Since
\begin{align}
\label{eq:kk00}
\sum_{k=0}^{2^{n-s_c-1}-1} \ket{k, k }_{AB}&=2^{\frac{n-s_c-1}{2}}\ket{0_00_0}_{AB}\otimes\ket{\bm{\phi}(n-1-s_c)},
\end{align}
\noindent and
\begin{align}
\label{eq:kk11}
\sum_{k=2^{n-s_c-1}}^{2^{n-s_c}-1} \ket{k, k }_{AB}&=2^{\frac{n-s_c-1}{2}}\ket{1_0 1_0}_{AB}\otimes\ket{\bm{\phi}(n-1-s_c)},
\end{align}
\noindent \cref{eq:chifinal0} reads
\begin{align}
\label{eq:chifinal}
&\ket{\chi}_{AB}
 = \frac{1}{2^n\sqrt{P(p_D)P(S)}}
 u\left(\delta_p\frac{\Delta}{\sqrt{2}}\right)
 \\ \nonumber
 & \times \left\{
 \begin{array}{ll}
\left[2^{\frac{n-s_c-1}{2}}\ket{0_00_0}_{AB}\otimes\ket{\bm{\phi}(n-1-s_c)}+\sum_{k=2^{n-s_c-1}}^{2^{n-s_c}-t_S-1} \ket{k, k}\right] \otimes \ket{S}_{AB} & \text{ if } t \in \left(0, 2^{n-1}\right]\\
2^{\frac{n-s_c-1}{2}}\left(\ket{0_00_0}_{AB}+\ket{1_0 1_0}_{AB}\right) \otimes\ket{\bm{\phi}(n-1-s_c)}\otimes\ket{S}_{AB}=2^{\frac{n-s_c}{2}}\ket{\bm{\phi}(n-s_c)} \otimes \ket{S}_{AB}&  \text{ if } t=0
\\
\left[\sum_{k=-t_S}^{2^{n-s_c-1}-1} \ket{k, k}+2^{\frac{n-s_c-1}{2}}\ket{1_0 1_0}_{AB}\otimes\ket{\bm{\phi}(n-s_c-1)}\right] \otimes \ket{S}_{AB} & \text{ if } t \in \left[ -2^{n-1},0 \right)
\end{array}
\right..
\end{align}

\Cref{eq:chifinal} shows that the outcome of the measurement of the first qubit should be correlated with the outcome of the homodyne measurement for a successful protocol. Thus, when the homodyne measurement yields a value of $p_D$ that implies $t \in \left(0, 2^{n-1}\right]$, the protocol is considered successful if the outcome of the first qubit measurement is $0$ at both Alice's and Bob's locations. For $t \in \left[-2^{n-1},0 \right)$, the protocol is successful if the first qubit measurement yields $1$ at both Alice's and Bob's locations. When $t=0$, the protocol is successful if the outcome of first qubit measurement at Alice's and Bob's is the same, either both $0$ or both $1$.

To summarize, the purification step for a protocol with Gaussian CV modes consists of the measurement of qubit $0$ and qubits
$n-s_c, \dots, n-1$ of Alice's and Bob's registers. The protocol is successful when
both propositions  {\it{i)}} and {\it{ii)}} are simultaneously true, where the proposition
{\it{i)}} is ``the outcome of the measurement of the $1+s_c$ qubits at Alice is the same as
the outcome at Bob'' and the proposition {\it{ii)}} is  ``the first qubit is $0$ and $t(p_d) \in \left[0, 2^{n-1}\right]$,  or, the first qubit is $1$ and $t(p_d) \in \left[-2^{n-1},0 \right]$ ''.

\subsubsection{Success probability for high-fidelity protocols}
\label{ssec:probhf}

The infidelity of our protocol is of the order of the truncation error introduced by~\cref{eq:ucutoff}. For sufficiently large $s_c$, these truncation errors  are negligible and the protocol yields $n_b$ Bell pairs with high-fidelity when the measurement outcomes satisfy the conditions summarized at the end of~\cref{ssec:pstep}.
Employing~\cref{eq:chifinal}, the conditional probability to produce
$n_b$ Bell pairs for a successful measurement outcome is:
\begin{align}
\label{eq:prbcond}
&P\left(\bm{\phi}(n-s_c-1) \bigm\vert j_{0,A},j_{n-s_c,A},\dots, j_{n-1,A}=
j_{0,B},j_{n-s_c,B},\dots, j_{n-1,B}=\{0,S\},p_D \in \left(-\frac{\Delta}{2\sqrt{2}}, \left(2^n+1\right)\frac{\Delta}{2\sqrt{2}} \right]\right)=\\ \nonumber
&P\left(\bm{\phi}(n-s_c-1)\bigm\vert j_{0,A},j_{n-s_c,A},\dots, j_{n-1,A}=
j_{0,B},j_{n-s_c,B},\dots, j_{n-1,B}=\{1,S\},p_D\in \left(-\left(2^n+1\right)\frac{\Delta}{2\sqrt{2}}, \frac{\Delta}{2\sqrt{2}} \right]\right)=\\ \nonumber
&=\frac{1}{2^{n+1+s_c}P(p_D)P(S)}
u\left[\delta_p\left(p_D\right)\frac{\Delta}{\sqrt{2}}\right]^2,
\end{align}
\noindent where $j_{q,A}$ ($j_{q,B}$) denotes the outcome of the measurement of qubit $q$ in
Alice's (Bob's) register and $S$ is defined by~\cref{eq:Sdef}.

Consequently, the protocol probability of producing $n_b$  Bell pairs is given by
\begin{align}
\label{eq:probnm1sc}
 P\left[\bm{\phi}(n-s_c-1)\right]&=\int_{-\left(2^{n-1}+1\right)\frac{\Delta}{2\sqrt{2}}}^{\frac{\Delta}{2\sqrt{2}}} P(p_D) \sum_S P(S)\frac{1}{2^{n+1+s_c}P(p_D)P(S)}
 u\left[\delta_p\left(p_D\right)\frac{\Delta}{\sqrt{2}}\right]^2 dp_D
 \\ \nonumber
 &+\int_{-\frac{\Delta}{2\sqrt{2}}}^{\left(2^{n-1}+1\right)\frac{\Delta}{2\sqrt{2}}} P(p_D)  \sum_S P(S)\frac{1}{2^{n+1+s_c}P(p_D)P(S)}
 u\left[\delta_p\left(p_D\right)\frac{\Delta}{\sqrt{2}}\right]^2 dp_D
 \\ \nonumber
 &=\frac{1}{2^{n+1}} \left(\sum_{t=-2^{n-1}}^{-1}+2+\sum_{t=1}^{2^{n-1}}\right)
\int_{-\frac{\Delta}{2\sqrt{2}}}^{\frac{\Delta}{2\sqrt{2}}} \abs{u\left(p\right)}^2 dp
 =\left(\frac{1}{2}+\frac{1}{2^n}\right) \erf\left(\frac{\sigma}{2} \sqrt{\frac{\pi}{2^n}}\right).
\end{align}
To derive~\cref{eq:probnm1sc} we used $\sum_S=2^{s_c}$ and
\begin{align}
\int_{-\frac{\Delta}{2\sqrt{2}}}^{\frac{\Delta}{2\sqrt{2}}}\abs{u\left(p\right)}^2 dp =\frac{\sigma}{\sqrt{\pi}}\int_{-\frac{\Delta}{2\sqrt{2}}}^{\frac{\Delta}{2\sqrt{2}}} dp e^{-\sigma^2 p^2}
%=\erf\left(\frac{\sigma\Delta}{2\sqrt{2}}\right)
=\erf\left(\frac{\sigma}{2} \sqrt{\frac{\pi}{2^n}}\right),
\end{align}
\noindent where $\erf()$ is the error function.

By inspecting~\cref{eq:chifinal}, one can see that when the homodyne measurement yields $t=0$, the resulting state  consists of $n-s_c$ Bell pairs, {\emph{i.e.}} the protocol produces one extra Bell pair compared to the case $t \ne 0$. The probability of our protocol  producing $n-s_c$ Bell pairs is
\begin{align}
\label{eq:probnmsc}
 P\left[\bm{\phi}(n-s_c)\right]&=\int_{-\frac{\Delta}{2\sqrt{2}}}^{\frac{\Delta}{2\sqrt{2}}} P(p_D) \sum_S P(S)\frac{1}{2^{n+s_c}P(p_D)P(S)}
 \abs{u\left[\delta_p\left(p_D\right)\frac{\Delta}{\sqrt{2}}\right]}^2 dp_D
 =\frac{1}{2^n} \erf\left(\frac{\sigma}{2} \sqrt{\frac{\pi}{2^n}}\right).
\end{align}
\noindent  Since having $n-s_c$ Bell pairs
automatically implies having $n-s_c-1$ Bell pairs, a term equal to $P\left[\bm{\phi}(n-s_c)\right]$
is also included in $P\left[\bm{\phi}(n-s_c-1)\right]$, as can be seen in~\cref{eq:probnm1sc}.

The probability to create $n-s_c$ Bell pairs is small for large $n$.
However, for small $n$, this probability has a significant
contribution to the success probability of the protocol.
For example, when  $n=2$, $s_c=0$ and the squeezing is large enough,  this  contribution is
$\approx 0.25$, bringing the
success probability for creating a single Bell pair up to $\approx 0.75$
(see discussion in~\cref{ssec:fidprob}).

\subsubsection{Protocol efficiency and fidelity}
\label{ssec:fidprob}

In this section, we investigate the success probability and fidelity of the protocol and calculate the optimal value of  $s_c$. The protocol is considered successful if fidelity is greater than a desired threshold. According to this definition, the success probability is a function of the fidelity threshold.

The purification step described in~\cref{ssec:pstep} was justified for
a choice of $s_c$ large enough  to ensure that the truncation errors in~\cref{eq:ucutoff} are small. This condition is met  when $s_c$ satisfies~\cref{eq:sigmascut}, which implies that
\begin{align}
\label{eq:scvssigma}
s_c \ge \tilde{s}(\sigma, n_b)=2 \log_2\left(\frac{c}{\sigma}\sqrt{\frac{2^{n_b}}{\pi}}+\sqrt{\frac{ 2^{n_b} c^2}{\pi \sigma^2}+1}\right)-1.
\end{align}

When the truncation errors are negligibly small,
the success probability of a protocol with  $\sigma$, $n_b$ and $s_c$ is given by
(see ~\cref{eq:probnm1sc}),
\begin{align}
\label{eq:prbsigmascnb}
P(\sigma, s_c, n_b) & = \left(\frac{1}{2}+\frac{1}{2^{n_b}2^{s_c+1}}\right) \erf\left(\frac{\sigma}{\sqrt{2^{s_c+1}}} \sqrt{\frac{\pi}{2^{n_b+2}}}\right),
\end{align}
\noindent and is monotonically decreasing with increasing $s_c$. This implies that the maximum probability to  produce $n_b$ Bell pairs at a fixed $\sigma$ is attained for the minimum integer $s_c$ satisfying~\cref{eq:scvssigma}, {\emph{i.e.}} when
\begin{align}
\label{eq:scutcrt}
s_c(\sigma, n_b)=\ceil{2 \log_2\left(\frac{c}{\sigma}\sqrt{\frac{2^{n_b}}{\pi}}+\sqrt{\frac{ 2^{n_b} c^2}{\pi \sigma^2}+1}\right)}-1.
\end{align}
\noindent Since $s_c(\sigma, n_b)$ takes integer values, it exhibits a
descending stair-step behavior as a function of increasing $\sigma$, jumping from the value $s_c$ to $s_c-1$ when
\begin{align}
\label{eq:sigmac}
\sigma=\sigma_c \equiv \frac{2c}{\left(2^{s_c+1}-1\right)}\sqrt{\frac{2^{n_b+s_c+1}}{\pi}}.
\end{align}
The maximum success probability is obtained by replacing $s_c$ in~\cref{eq:prbsigmascnb}
with $s_c(\sigma, n_b)$ given by~\cref{eq:scutcrt},
\begin{align}
\label{eq:Pmax}
P(\sigma, n_b)=P(\sigma, s_c, n_b)|_{s_c=s_c(\sigma, n_b)}.
\end{align}
\noindent
The discontinuity of  $s_c(\sigma, n_b)$  manifests as a discontinuity in  $P(\sigma, n_b)$  as a function of  $\sigma$,  for values of  $\sigma$  that satisfy~\cref{eq:sigmac}. Notice that $s_c(\sigma, n_b)$ and $P(\sigma, n_b)$ depend implicitly on the fidelity threshold since they depend on the cutoff parameter $c$, and $c$ exerts strong influence on the fidelity.

To analyze the efficiency of our protocol for large $n_b$
it is useful to determine analytical bounds for the maximum success probability,
\begin{align}
\label{eq:Pbounds}
P_{lower} \left(\sigma, n_b \right) < P\left(\sigma, n_b \right) \le P_{upper}\left(\sigma, n_b \right).
\end{align}
Employing~\cref{eq:scvssigma} we find
\begin{align}
\label{eq:Pupbound}
P_{upper}\left(\sigma, n_b \right)= \left(\frac{1}{2}+\frac{1}{2^{n_b}}\frac{\pi \sigma^2}{\left(c\sqrt{2^{n_b}}+\sqrt{ 2^{n_b} c^2+\pi \sigma^2}\right)^2}\right)\erf\left(\frac{\pi\sigma^2 }{ 2 \sqrt{2^{n_b}}\left( c \sqrt{2^{n_b}} +\sqrt{2^{n_b} c^2+\pi \sigma^2}\right)}\right),
\end{align}
\noindent while the relation $s_c<\tilde{s}(\sigma, n_b)+1$ (implied by \cref{eq:scvssigma,eq:scutcrt}) yields
\begin{align}
\label{eq:Pdwbound}
P_{lower} \left(\sigma, n_b \right)= \left(\frac{1}{2}+\frac{1}{2^{n_b+1}}\frac{\pi \sigma^2}{\left(c\sqrt{2^{n_b}}+\sqrt{ 2^{n_b} c^2+\pi \sigma^2}\right)^2}\right)\erf\left(\frac{\pi\sigma^2 }{ 2 \sqrt{2^{n_b+1}}\left( c \sqrt{2^{n_b}} +\sqrt{2^{n_b} c^2+\pi \sigma^2}\right)}\right).
\end{align}

\begin{figure}[tb]
    \begin{center}
        \includegraphics*[width=5in]{sigmavsnbc2p2.eps}
        \caption{ The squeezing  parameter  $\ts(dB)$ necessary to produce $n_b$ Bell pairs
        for different target probabilities of success, plotted against $n_b$.
        The calculations use a cutoff parameter $c=2.2$, ensuring a fidelity $F > 0.99$. The dependence is linear for large $n_b$, with a slope of approximately $3.01 dB$ per Bell pair, as predicted by~\cref{eq:sigmavsnbPtg}. }
        \label{fig:sigmavsnb}
    \end{center}
\end{figure}

As can be inferred from~\cref{eq:probnm1sc}, the success probability increases with $\sigma/\sqrt{2^n}$, reaching the maximum value  of $\approx 0.5 + 0.5^n$ for large squeezing factors ({\emph{i.e.} large $\sigma$}). However, squeezing is a technologically expensive resource, with the maximum squeezing achieved with current technology being $15 dB$~\cite{Vahlbruch_prl_2016}. The present technological limitation of the squeezing factor restricts the number of Bell pairs that can be produced by our protocol.

For a fixed success probability, the minimum squeezing necessary to produce  $n_b$  Bell pairs, expressed in decibels, is proportional to  $n_b$, with a slope of approximately $3.01 dB$ per Bell pair. The minimum squeezing necessary to produce  $n_b$  Bell pairs can be calculated by solving  $P(\sigma, n_b) = P_{tg}$, where  $P_{tg}$  is the target success probability and  $P(\sigma, n_b)$  is given by~\cref{eq:Pmax}. The solutions, obtained numerically, are illustrated in~\cref{fig:sigmavsnb}, where $\ts(dB)$ is defined as:
\begin{align}
\label{eq:sigmadecib}
\ts(dB) \equiv 20 \log_{10} (\sigma).
\end{align}
The linear dependence of $\ts$  versus  $n_b$  can be understood by inspecting \cref{eq:Pupbound,eq:Pdwbound}. Ignoring the term proportional to  $\frac{1}{2^{n_b}}$  (which becomes small for large  $n$ ), we can approximate:
\begin{align}
P_{upper,lower} \left(\sqrt{2}\sigma, n_b+1\right) \approx P_{upper,lower} \left(\sigma, n_b\right).
\end{align}

This implies, following an induction argument, that
$P\left(\frac{\sigma}{\sqrt{2^{n_b}}}, 0\right) \approx P_{tg}$, which in turn implies that
$\frac{\sigma}{\sqrt{2^{n_b}}} \approx C(P_{tg})$, where  $C(P_{tg})$  is a function that depends on the target success probability. We can write:
\begin{align}
\label{eq:sigmavsnbPtg}
\ts(dB) & \approx 10 \log_{10}(2) \times n_b + 20 \log_{10} \left[C(P_{tg})\right] \approx 3.01  dB \times n_b + C^{\prime}(P_{tg}),
\end{align}
which agrees with the plots presented in \cref{fig:sigmavsnb}.

In particular, the squeezing necessary for the maximum success probability can be estimated
analytically. The required condition is that $\erf\left(\frac{\sigma}{2} \sqrt{\frac{\pi}{2^n}}\right) \approx 1$ (see~\cref{eq:probnm1sc}), which implies $\sigma \ge 2c\sqrt{\frac{2^n}{\pi}}$ and $s_c=0$ (see~\cref{eq:sigmamcut}). Employing
~\cref{eq:probnm1sc,eq:sigmadecib}, we conclude
\begin{align}
\label{eq:sigmadecibnm1}
P(\sigma, n-1)\approx 0.5+0.5^n,~\text{   when   }~
\ts(dB) \ge 10 \log_{10}\left(2\right) \times n+ 20 \log_{10}\left(\frac{2c}{\sqrt{\pi},
}\right) \approx 3.01 dB \times n+ C. %7.9 dB,
 \end{align}
\noindent
Here, $C$ is a term which depends on the threshold fidelity, but is independent of $n$. For example, for $c=2.2$ which yields
a fidelity $F>0.99$, we obtain $C=7.9dB$.

\begin{figure}[tb]
    \begin{center}
        \includegraphics*[width=5in]{Pvssigma_nb3_v1.eps}
        \caption{The protocol success probability versus squeezing $\ts(dB)$.
        The symbols are obtained  by simulating the protocol steps outlined in~\cref{sec:es_protocol} for a protocol producing $n_b=3$ Bell states with fidelity $F>0.99$, for different values of $s_c$. Since the truncation approximation (\cref{eq:ucutoff}) fails when $\sigma<\sigma_c$ (see~\cref{eq:sigmascut,eq:sigmac}),
         the probability of obtaining high-fidelity Bell states rapidly drops
        to zero when $\sigma < \sigma_c$,  while keeping  $s_c$  fixed.
        The dotted lines correspond to  $P(\sigma, s_c, n_b)$ given by~\cref{eq:prbsigmascnb} for  $\sigma \ge \sigma_c$  and to zero for  $\sigma < \sigma_c$.
        The dashed lines represent $P_{upper}(\sigma,  n_b)$ and $P_{lower}(\sigma,  n_b)$ given by~\cref{eq:Pupbound,eq:Pdwbound}, and the solid line represents the  expression for the success probability given by~\cref{eq:Pmax}.
        When  $\sigma \approx \sigma_c$, the maximum success probability jumps up because it corresponds to the success probability obtained when  $s_c$  is reduced by $1$.
        All analytical results are calculated with $c=2.17$, a value that best fits the numerical results for a threshold fidelity of $0.99$. }
         \label{fig:Pvssigma}
    \end{center}
\end{figure}

For high-fidelity thresholds, the analytical results presented so far are in agreement with numerical simulations. This is illustrated in~\cref{fig:Pvssigma} which plots the success probability versus squeezing $\ts(\text{dB})$ for a protocol producing $n_b = 3$ Bell pairs with a fidelity exceeding $0.99$, for different values of $s_c$.
The success probability is obtained numerically by simulating the protocol steps outlined at the beginning of~\cref{sec:es_protocol} and
counting all the final states that pass the purification rules summarized at the end of~\cref{ssec:pstep} and have a fidelity exceeding the threshold fidelity.
The analytical approximation of the success probability,  $P(\sigma, n_b)$, given by~\cref{eq:Pmax}, is shown with a continuous line, while  $P_{upper}(\sigma, n_b)$  and  $P_{lower}(\sigma, n_b)$, described by~\cref{eq:Pupbound,eq:Pdwbound}, are shown with dashed lines.
Notice the discontinuity of  $P(\sigma, n_b)$, which jumps from  $P_{lower}(\sigma, n_b)$  to  $P_{upper}(\sigma, n_b)$  for values of  $\sigma$  that satisfy~\cref{eq:sigmac}, a direct consequence of the discontinuity of  $s_c(\sigma, n_b)$  as a function of  $\sigma$ (see~\cref{eq:scutcrt}).
At these points, $\sigma$  is large enough that a decrease in  $s_c$  by $1$, while maintaining the high-fidelity approximation (\cref{eq:ucutoff}), is possible. The decrease in $s_c$ results in a jump in the success probability.
On the intervals between the discontinuity points,  $P(\sigma, n_b)$ is described by~\cref{eq:prbsigmascnb}.
The best fit between analytical predictions and numerical calculations with $F=0.99$ fidelity
threshold is obtained by taking the cutoff parameter $c = 2.17$
in~\cref{eq:scvssigma,eq:Pmax,eq:Pupbound,eq:Pdwbound}. The good agreement validates the analytical approach and indicates that for a cutoff parameter $c \gtrapprox 2.17$, the truncation errors introduced by~\cref{eq:ucutoff} are small enough to yield fidelity errors of $\approx 1\%$.

% Notice that for $s_c = 0$ (open triangles) there are discrepancies due to truncation errors in the approximation~\cref{eq:ucutoff}, which persist even for the chosen fidelity threshold of $0.99$. As expected, simulations with a higher fidelity requirement (not shown) exhibit even better agreement with the analytical predictions.

\begin{figure}[tb]
    \begin{center}
        \includegraphics*[width=5in]{prbvsF.eps}
        \caption{ Success probability versus threshold fidelity
        for a protocol producing $n_b=4$ Bell pairs, for  a) $\ts=10dB$ squeezing and
        b) $\ts=15dB$ squeezing.
        The success probability increases as the threshold fidelity decreases.
        When $s_c$ is large enough to satisfy~\cref{eq:sigmascut}  for a value of
        $c \gtrapprox 2.2$ (green diamond in (b) and blue
        triangle in (a) and (b)), the success probability is given by~\cref{eq:prbsigmascnb}
        up to a threshold fidelity $>0.99$.}
        \label{fig:prbvsF}
    \end{center}
\end{figure}

When the value of $s_c$ is not large enough to make the truncation errors in~\cref{eq:ucutoff} negligible, the fidelity of the protocol is reduced. Numerical calculations of
the success probability versus the threshold fidelity are illustrated in~\cref{fig:prbvsF} for different values of $s_c$ for a protocol generating $n_b=4$ Bell pairs.
The success probability reaches its maximum  when all states passing the purification rules are counted.
We identify three regimes for qualitative distinction.
In the low squeezing and small $s_c$ regime (for example $s_c=0$ and  $s_c=1$ at $\ts=10dB$ in~\cref{fig:prbvsF} -a), the fidelity of all the states is low, and the protocol is essentially failing.
In the intermediate $s_c$ and squeezing regime, the success probability is small for a protocol with high fidelity threshold (for example, $F > 0.9$), since many states passing the purification rules have low fidelity. Decreasing the threshold fidelity significantly increases the success probability in this case (for example, observe $s_c=2$ at $\ts=10 \text{ dB}$ in~\cref{fig:prbvsF} -a and $s_c=0$ and $s_c=1$ at $\ts=15 \text{ dB}$ in~\cref{fig:prbvsF} -b).
Finally, we have the high-fidelity regime, achieved with large $s_c$ and/or squeezing (see $s_c=3$ at $\ts=10 \text{ dB}$ in~\cref{fig:prbvsF} -a and $s_c \ge 2$ at $\ts=15 \text{ dB}$ in~\cref{fig:prbvsF} -b). In this regime, all states passing the purification rules have high fidelity ($F>0.99$). The success probability is well approximated by~\cref{eq:prbsigmascnb}.

\begin{figure}[tb]
    \begin{center}
        \includegraphics*[width=5in]{prbvsFnb1.eps}
        \caption{ Success probability versus threshold fidelity for a protocol producing one Bell pair.  For a threshold fidelity $F>0.8$, the success probability is significantly larger than $0.5$ for squeezing values achievable with present-day technology.}
        \label{fig:prbvsFnb1}
    \end{center}
\end{figure}

Although the main advantage of our protocol is the capability to produce  multiple Bell pairs, it can also be used to produce single Bell pairs (by taking $n=2$ and $s_c=0$) with success probabilities reaching $0.75$ for squeezing achievable with present technology (i.e., $\ts \le 15 dB$). This is predicted by~\cref{eq:sigmadecibnm1} when $\ts \gtrapprox 14 dB$ and by numerical calculations shown in~\cref{fig:prbvsFnb1}, where the success probability versus fidelity is plotted for different values of the squeezing factor. For example, when $\ts = 10dB$ ($\ts = 15 dB$), our protocol can produce a Bell pair with $0.9$ ($0.99$) fidelity with a success probability of approximately $0.7$ ($0.75$). These success probabilities are significantly larger than the value of $0.5$, which represents the upper bound characterizing photonic-qubit protocols with incomplete BSM~\cite{Calsamiglia_apb_2001}, giving our protocol a potential advantage for generating Bell pairs at a higher rate when employed in optical networks.

\section{Discussions}
\label{sec:disc}

Several potential impediments to the practical implementation of the protocol in the near future need to be addressed. For example, step 6 of our protocol requires the implementation of the multi-qubit operator $D(-t)$ (see~\cref{eq:Pdisp}) at one of the nodes. This necessitates the implementation of two multi-qubit QFTs (see~\cref{fig:exptP}), which requires $\O(n^2)$ two-qubit gates~\cite{nielsen2010quantum} (such as CNOT gates). For large-sized registers, this can be challenging since the limited fidelity of two-qubit gates with current technology restricts the number of two-qubit gates that can be implemented.
However the implementation for small $n$ is feasible  with current technology. For example, the QFT for $n=2$ requires only one CNOT gate.

In our protocol, the number of Bell pairs that can be produced increases with increasing squeezing, as implied by~\cref{eq:sigmadecibnm1}. The current technological limitations in squeezing restrict the protocol’s feasibility to generating only a small number of Bell pairs at present. However, significant efforts to advance squeezing technology are underway in many groups~\cite{Schonbeck_2018,Vahlbruch_prl_2016,Hagemann_2024,Li_2022}, enhancing the relevance of our protocol for near-future applications.

While we considered squeezed vacuum here, the protocol also works with other choices of the initial CV wavefunction. The essential requirement is that the function $\hh_{x_D}(p)$, defined in~\cref{eq:hhfun}, has a narrow support centered around $p=0$ within a good approximation.
For instance, we implemented the protocol using a rectangular $g(x)$, although we did not conduct a comprehensive investigation of its efficiency in this context. Exploring alternative choices could prove beneficial for optimizing the protocol’s practical implementation under varying technological constraints.

An essential goal of any entanglement generation protocol is to generate high-fidelity Bell pairs at a high rate for quantum networks.
For example, a direct application of our protocol would be a quantum repeater designed to distribute Bell pairs between multi-qubit end nodes separated by long distances. Between the end nodes, a series of intermediary nodes will be placed, each containing two registers of  $n$  qubits. Midway between these qubit nodes, optical nodes equipped with beam splitters and homodyne detectors are positioned. These optical nodes are connected via fiber optics to the adjacent qubit nodes. The maximum distance between the intermediary nodes is constrained by tolerable loss in fiber optics, which will be estimated in future work.
Our protocol constitutes the first step in the long-distance entanglement distribution process, generating  $n_b$  Bell pairs between adjacent qubit nodes by employing beam splitters and homodyne measurements at the optical nodes. The next step will involve local and deterministic qubit Bell measurements at the intermediary nodes, progressively extending the range of entanglement. The final outcome will be multiple Bell pairs shared between the end nodes.
Various purification protocols may also need to be considered in realistic scenarios to compensate for noise-induced errors.

Since the CV-DV hybrid protocol presented here yields multiple entangled qubits, it is worth evaluating its efficiency against other protocols capable of producing multiple Bell pairs.
For example, qubit-based DV-DV protocols can generate at most a single Bell pair per swapping event. To produce multiple Bell pairs, they must be applied multiple times. The efficiency of DV-DV protocols in generating multiple Bell pairs depends on the employed signal transmission scheme (multiplexing), success probability, qubit coherence time, signal loss, and distillation procedures.

Other hybrid CV-DV protocols capable of generating multiple Bell pairs should also be considered for comparison. For instance, the protocol introduced in Ref.~\cite{vanLoock_PRL_2006}, which employs a dispersive light-matter interaction between a coherent CV mode and qubits, could be generalized. By controlling the interaction strength or duration, the phase of the coherent mode can be made dependent on the number  $j$  that defines the DV state (see \cref{eq:jbasis}). We envision two possible approaches for the next step. First, two distinct coherent modes could be used: one interacting with the DV register at Alice and the other with the DV register at Bob, after which both modes would be sent to Charlie and passed through a beam splitter. Alternatively, a single CV mode could interact with both DV registers. In both cases, homodyne measurements on the CV states (or the single CV state in the latter case) could, in principle, project,  with a finite success probability, a state that has a large overlap with a multi-Bell-pair state.
As in the protocol introduced in Ref.~\cite{vanLoock_PRL_2006}, achieving high fidelity will likely require large-amplitude coherent states, with the amplitude increasing as  $n$  increases. Beam loss is expected to impose significant constraints on the protocol’s success probability.

Many factors influence the Bell pair generation rate, including quantum and classical signal loss in optical fibers, homodyne and qubit measurement errors, distillation protocols, memory decoherence, and signal multiplexing and parallelization schemes. Our protocol, capable of generating both multiple and single Bell pairs with high success probability, holds promise for increasing the Bell pair generation rate.
However, a quantitative estimate can only be made after integrating the protocol into an optimized quantum repeater or networking framework while accounting for loss and errors.
For example, by introducing fictitious beam splitters coupled to vacuum modes, one can model the effects of beam loss and beam mismatch in fiber-optic setups and homodyne detectors. Phase noise in homodyne detection can be accounted for by introducing quadrature phase shifts. Electronic noise in photodiodes can be represented by adding a small classical error to the measured quadrature values. We plan to investigate these effects on our protocol in future work.

\section{Conclusions}
\label{sec:conc}

We present a hybrid DV-CV entanglement generation protocol using linear optics and homodyne measurements, capable of producing multiple Bell pairs per protocol iteration with  $\approx 0.5$ probability.
Two multi-qubit registers of size  $n$  at two separate nodes are locally entangled with squeezed vacuum optical modes. The optical modes are sent to a third party, passed through a $50:50$ beam-splitter, and undergo homodyne quadrature measurement. The outcome of the homodyne measurement is transmitted back to the multi-qubit nodes through classical channels. Local operations, such as single qubit phase rotations with angles determined by the homodyne measurement outcome and QFT, are then applied to the multi-qubit registers, followed by the measurement of $1+s_c$ qubits at each node. The success of the protocol depends on the outcomes of the homodyne and qubit measurements. When successful, the protocol produces $n_b=n-1-s_c$ qubits with high fidelity.

The efficiency of our protocol is limited by the technological constraints in squeezing optical modes. With sufficiently large squeezing (see~\cref{eq:sigmadecibnm1}), only one qubit at each register needs to be measured, and the protocol’s success probability for producing $n-1$ Bell pairs with a fidelity exceeding $0.99$ reaches $0.5 + 0.5^n$. The success probability and fidelity decreases with decreasing squeezing strength. The fidelity can be increased by measuring more qubits ({\emph{i.e.}} by increasing $s_c$), but this decreases both the success probability and the final number of Bell pairs. To maintain a fixed success probability, an additional squeezing of $3.01 dB$ is required for every extra Bell pair produced.

Our protocol can also be used to produce high-fidelity single Bell pairs with a success probability reaching $0.75$ for squeezing strengths achievable with current technology.

Both the capability to generate multiple Bell pairs and the ability to produce a single Bell pair with a high success probability demonstrate the potential of our protocol to advance the development of new quantum repeaters and quantum networks with high-rate Bell pair generation.

Our protocol can generate multiple Bell pairs even with small squeezing, easily achievable with current experimental techniques, albeit with a low success probability that limits its effectiveness. However, this opens up the possibility for experimental testing.

\section{Acknowledgments}

This manuscript has been authored by Fermi Research Alliance, LLC under Contract No.
DE-AC02-07CH11359 with the U.S. Department of Energy, Office of Science, Office of High Energy Physics.
The research team are partially supported by the U.S. Department of Energy, Office of Science Advanced Scientific Computing Research program under FWP FNAL
23-24.
%\bibliography{bibmacridin}

\begin{thebibliography}{29}%
\makeatletter
\providecommand \@ifxundefined [1]{%
 \@ifx{#1\undefined}
}%
\providecommand \@ifnum [1]{%
 \ifnum #1\expandafter \@firstoftwo
 \else \expandafter \@secondoftwo
 \fi
}%
\providecommand \@ifx [1]{%
 \ifx #1\expandafter \@firstoftwo
 \else \expandafter \@secondoftwo
 \fi
}%
\providecommand \natexlab [1]{#1}%
\providecommand \enquote  [1]{``#1''}%
\providecommand \bibnamefont  [1]{#1}%
\providecommand \bibfnamefont [1]{#1}%
\providecommand \citenamefont [1]{#1}%
\providecommand \href@noop [0]{\@secondoftwo}%
\providecommand \href [0]{\begingroup \@sanitize@url \@href}%
\providecommand \@href[1]{\@@startlink{#1}\@@href}%
\providecommand \@@href[1]{\endgroup#1\@@endlink}%
\providecommand \@sanitize@url [0]{\catcode `\\12\catcode `\$12\catcode
  `\&12\catcode `\#12\catcode `\^12\catcode `\_12\catcode `\%12\relax}%
\providecommand \@@startlink[1]{}%
\providecommand \@@endlink[0]{}%
\providecommand \url  [0]{\begingroup\@sanitize@url \@url }%
\providecommand \@url [1]{\endgroup\@href {#1}{\urlprefix }}%
\providecommand \urlprefix  [0]{URL }%
\providecommand \Eprint [0]{\href }%
\providecommand \doibase [0]{https://doi.org/}%
\providecommand \selectlanguage [0]{\@gobble}%
\providecommand \bibinfo  [0]{\@secondoftwo}%
\providecommand \bibfield  [0]{\@secondoftwo}%
\providecommand \translation [1]{[#1]}%
\providecommand \BibitemOpen [0]{}%
\providecommand \bibitemStop [0]{}%
\providecommand \bibitemNoStop [0]{.\EOS\space}%
\providecommand \EOS [0]{\spacefactor3000\relax}%
\providecommand \BibitemShut  [1]{\csname bibitem#1\endcsname}%
\let\auto@bib@innerbib\@empty
%</preamble>
\bibitem [{\citenamefont {\ifmmode~\dot{Z}\else \.{Z}\fi{}ukowski}\ \emph
  {et~al.}(1993)\citenamefont {\ifmmode~\dot{Z}\else \.{Z}\fi{}ukowski},
  \citenamefont {Zeilinger}, \citenamefont {Horne},\ and\ \citenamefont
  {Ekert}}]{zukowski_prl_1993}%
  \BibitemOpen
  \bibfield  {author} {\bibinfo {author} {\bibfnamefont {M.}~\bibnamefont
  {\ifmmode~\dot{Z}\else \.{Z}\fi{}ukowski}}, \bibinfo {author} {\bibfnamefont
  {A.}~\bibnamefont {Zeilinger}}, \bibinfo {author} {\bibfnamefont {M.~A.}\
  \bibnamefont {Horne}},\ and\ \bibinfo {author} {\bibfnamefont {A.~K.}\
  \bibnamefont {Ekert}},\ }\bibfield  {title} {\bibinfo {title}
  {{``E}vent-ready-detectors{''} {B}ell experiment via entanglement swapping},\
  }\href {https://doi.org/10.1103/PhysRevLett.71.4287} {\bibfield  {journal}
  {\bibinfo  {journal} {Phys. Rev. Lett.}\ }\textbf {\bibinfo {volume} {71}},\
  \bibinfo {pages} {4287} (\bibinfo {year} {1993})}\BibitemShut {NoStop}%
\bibitem [{\citenamefont {Pan}\ \emph {et~al.}(1998)\citenamefont {Pan},
  \citenamefont {Bouwmeester}, \citenamefont {Weinfurter},\ and\ \citenamefont
  {Zeilinger}}]{pan_prl_1998}%
  \BibitemOpen
  \bibfield  {author} {\bibinfo {author} {\bibfnamefont {J.-W.}\ \bibnamefont
  {Pan}}, \bibinfo {author} {\bibfnamefont {D.}~\bibnamefont {Bouwmeester}},
  \bibinfo {author} {\bibfnamefont {H.}~\bibnamefont {Weinfurter}},\ and\
  \bibinfo {author} {\bibfnamefont {A.}~\bibnamefont {Zeilinger}},\ }\bibfield
  {title} {\bibinfo {title} {Experimental entanglement swapping: Entangling
  photons that never interacted},\ }\href
  {https://doi.org/10.1103/PhysRevLett.80.3891} {\bibfield  {journal} {\bibinfo
   {journal} {Phys. Rev. Lett.}\ }\textbf {\bibinfo {volume} {80}},\ \bibinfo
  {pages} {3891} (\bibinfo {year} {1998})}\BibitemShut {NoStop}%
\bibitem [{\citenamefont {van Loock}\ and\ \citenamefont
  {Braunstein}(1999)}]{Look_PRA_1999}%
  \BibitemOpen
  \bibfield  {author} {\bibinfo {author} {\bibfnamefont {P.}~\bibnamefont {van
  Loock}}\ and\ \bibinfo {author} {\bibfnamefont {S.~L.}\ \bibnamefont
  {Braunstein}},\ }\bibfield  {title} {\bibinfo {title} {Unconditional
  teleportation of continuous-variable entanglement},\ }\href
  {https://doi.org/10.1103/PhysRevA.61.010302} {\bibfield  {journal} {\bibinfo
  {journal} {Phys. Rev. A}\ }\textbf {\bibinfo {volume} {61}},\ \bibinfo
  {pages} {010302} (\bibinfo {year} {1999})}\BibitemShut {NoStop}%
\bibitem [{\citenamefont {L\"utkenhaus}\ \emph {et~al.}(1999)\citenamefont
  {L\"utkenhaus}, \citenamefont {Calsamiglia},\ and\ \citenamefont
  {Suominen}}]{Lutkenhaus_pra_1999}%
  \BibitemOpen
  \bibfield  {author} {\bibinfo {author} {\bibfnamefont {N.}~\bibnamefont
  {L\"utkenhaus}}, \bibinfo {author} {\bibfnamefont {J.}~\bibnamefont
  {Calsamiglia}},\ and\ \bibinfo {author} {\bibfnamefont {K.-A.}\ \bibnamefont
  {Suominen}},\ }\bibfield  {title} {\bibinfo {title} {Bell measurements for
  teleportation},\ }\href {https://doi.org/10.1103/PhysRevA.59.3295} {\bibfield
   {journal} {\bibinfo  {journal} {Phys. Rev. A}\ }\textbf {\bibinfo {volume}
  {59}},\ \bibinfo {pages} {3295} (\bibinfo {year} {1999})}\BibitemShut
  {NoStop}%
\bibitem [{\citenamefont {Vaidman}\ and\ \citenamefont
  {Yoran}(1999)}]{Vaidman_PRA_1999}%
  \BibitemOpen
  \bibfield  {author} {\bibinfo {author} {\bibfnamefont {L.}~\bibnamefont
  {Vaidman}}\ and\ \bibinfo {author} {\bibfnamefont {N.}~\bibnamefont
  {Yoran}},\ }\bibfield  {title} {\bibinfo {title} {Methods for reliable
  teleportation},\ }\href {https://doi.org/10.1103/PhysRevA.59.116} {\bibfield
  {journal} {\bibinfo  {journal} {Phys. Rev. A}\ }\textbf {\bibinfo {volume}
  {59}},\ \bibinfo {pages} {116} (\bibinfo {year} {1999})}\BibitemShut
  {NoStop}%
\bibitem [{\citenamefont {Calsamiglia}\ and\ \citenamefont
  {L{\"u}tkenhaus}(2001)}]{Calsamiglia_apb_2001}%
  \BibitemOpen
  \bibfield  {author} {\bibinfo {author} {\bibfnamefont {J.}~\bibnamefont
  {Calsamiglia}}\ and\ \bibinfo {author} {\bibfnamefont {N.}~\bibnamefont
  {L{\"u}tkenhaus}},\ }\bibfield  {title} {\bibinfo {title} {Maximum efficiency
  of a linear-optical {B}ell-state analyzer},\ }\href@noop {} {\bibfield
  {journal} {\bibinfo  {journal} {Applied Physics B}\ }\textbf {\bibinfo
  {volume} {72}},\ \bibinfo {pages} {67} (\bibinfo {year} {2001})}\BibitemShut
  {NoStop}%
\bibitem [{\citenamefont {Kwiat}\ and\ \citenamefont
  {Weinfurter}(1998)}]{Kwiat_pra_1998}%
  \BibitemOpen
  \bibfield  {author} {\bibinfo {author} {\bibfnamefont {P.~G.}\ \bibnamefont
  {Kwiat}}\ and\ \bibinfo {author} {\bibfnamefont {H.}~\bibnamefont
  {Weinfurter}},\ }\bibfield  {title} {\bibinfo {title} {Embedded {B}ell-state
  analysis},\ }\href {https://doi.org/10.1103/PhysRevA.58.R2623} {\bibfield
  {journal} {\bibinfo  {journal} {Phys. Rev. A}\ }\textbf {\bibinfo {volume}
  {58}},\ \bibinfo {pages} {R2623} (\bibinfo {year} {1998})}\BibitemShut
  {NoStop}%
\bibitem [{\citenamefont {Kim}\ \emph {et~al.}(2001)\citenamefont {Kim},
  \citenamefont {Kulik},\ and\ \citenamefont {Shih}}]{kim_prl_2001}%
  \BibitemOpen
  \bibfield  {author} {\bibinfo {author} {\bibfnamefont {Y.-H.}\ \bibnamefont
  {Kim}}, \bibinfo {author} {\bibfnamefont {S.~P.}\ \bibnamefont {Kulik}},\
  and\ \bibinfo {author} {\bibfnamefont {Y.}~\bibnamefont {Shih}},\ }\bibfield
  {title} {\bibinfo {title} {Quantum teleportation of a polarization state with
  a complete {B}ell state measurement},\ }\href
  {https://doi.org/10.1103/PhysRevLett.86.1370} {\bibfield  {journal} {\bibinfo
   {journal} {Phys. Rev. Lett.}\ }\textbf {\bibinfo {volume} {86}},\ \bibinfo
  {pages} {1370} (\bibinfo {year} {2001})}\BibitemShut {NoStop}%
\bibitem [{\citenamefont {Barrett}\ \emph {et~al.}(2005)\citenamefont
  {Barrett}, \citenamefont {Kok}, \citenamefont {Nemoto}, \citenamefont
  {Beausoleil}, \citenamefont {Munro},\ and\ \citenamefont
  {Spiller}}]{Barrett_pra_2005}%
  \BibitemOpen
  \bibfield  {author} {\bibinfo {author} {\bibfnamefont {S.~D.}\ \bibnamefont
  {Barrett}}, \bibinfo {author} {\bibfnamefont {P.}~\bibnamefont {Kok}},
  \bibinfo {author} {\bibfnamefont {K.}~\bibnamefont {Nemoto}}, \bibinfo
  {author} {\bibfnamefont {R.~G.}\ \bibnamefont {Beausoleil}}, \bibinfo
  {author} {\bibfnamefont {W.~J.}\ \bibnamefont {Munro}},\ and\ \bibinfo
  {author} {\bibfnamefont {T.~P.}\ \bibnamefont {Spiller}},\ }\bibfield
  {title} {\bibinfo {title} {Symmetry analyzer for nondestructive {B}ell-state
  detection using weak nonlinearities},\ }\href
  {https://doi.org/10.1103/PhysRevA.71.060302} {\bibfield  {journal} {\bibinfo
  {journal} {Phys. Rev. A}\ }\textbf {\bibinfo {volume} {71}},\ \bibinfo
  {pages} {060302} (\bibinfo {year} {2005})}\BibitemShut {NoStop}%
\bibitem [{\citenamefont {Jian}\ \emph {et~al.}(2009)\citenamefont {Jian},
  \citenamefont {Ming}, \citenamefont {Yan},\ and\ \citenamefont
  {Zhuo-Liang}}]{Jian_cpl_2009}%
  \BibitemOpen
  \bibfield  {author} {\bibinfo {author} {\bibfnamefont {Z.}~\bibnamefont
  {Jian}}, \bibinfo {author} {\bibfnamefont {Y.}~\bibnamefont {Ming}}, \bibinfo
  {author} {\bibfnamefont {L.}~\bibnamefont {Yan}},\ and\ \bibinfo {author}
  {\bibfnamefont {C.}~\bibnamefont {Zhuo-Liang}},\ }\bibfield  {title}
  {\bibinfo {title} {Nearly deterministic teleportation of a photonic qubit
  with weak cross-{K}err nonlinearities},\ }\href
  {https://doi.org/10.1088/0256-307X/26/10/100301} {\bibfield  {journal}
  {\bibinfo  {journal} {Chinese Physics Letters}\ }\textbf {\bibinfo {volume}
  {26}},\ \bibinfo {pages} {100301} (\bibinfo {year} {2009})}\BibitemShut
  {NoStop}%
\bibitem [{\citenamefont {Bai}\ \emph {et~al.}(2011)\citenamefont {Bai},
  \citenamefont {Guo}, \citenamefont {Cheng}, \citenamefont {Shao},
  \citenamefont {Wang}, \citenamefont {Zhang},\ and\ \citenamefont
  {Yeon}}]{Bai_2011}%
  \BibitemOpen
  \bibfield  {author} {\bibinfo {author} {\bibfnamefont {J.}~\bibnamefont
  {Bai}}, \bibinfo {author} {\bibfnamefont {Q.}~\bibnamefont {Guo}}, \bibinfo
  {author} {\bibfnamefont {L.-Y.}\ \bibnamefont {Cheng}}, \bibinfo {author}
  {\bibfnamefont {X.-Q.}\ \bibnamefont {Shao}}, \bibinfo {author}
  {\bibfnamefont {H.-F.}\ \bibnamefont {Wang}}, \bibinfo {author}
  {\bibfnamefont {S.}~\bibnamefont {Zhang}},\ and\ \bibinfo {author}
  {\bibfnamefont {K.-H.}\ \bibnamefont {Yeon}},\ }\bibfield  {title} {\bibinfo
  {title} {Implementation of nonlocal {B}ell-state measurement and quantum
  information transfer with weak {K}err nonlinearity},\ }\href
  {https://doi.org/10.1088/1674-1056/20/12/120307} {\bibfield  {journal}
  {\bibinfo  {journal} {Chinese Physics B}\ }\textbf {\bibinfo {volume} {20}},\
  \bibinfo {pages} {120307} (\bibinfo {year} {2011})}\BibitemShut {NoStop}%
\bibitem [{\citenamefont {Wang}\ \emph {et~al.}(2017)\citenamefont {Wang},
  \citenamefont {Zhu}, \citenamefont {Wang},\ and\ \citenamefont
  {Ye}}]{Wang_QIP_2017}%
  \BibitemOpen
  \bibfield  {author} {\bibinfo {author} {\bibfnamefont {J.-M.}\ \bibnamefont
  {Wang}}, \bibinfo {author} {\bibfnamefont {M.-z.}\ \bibnamefont {Zhu}},
  \bibinfo {author} {\bibfnamefont {D.}~\bibnamefont {Wang}},\ and\ \bibinfo
  {author} {\bibfnamefont {L.}~\bibnamefont {Ye}},\ }\bibfield  {title}
  {\bibinfo {title} {Nearly deterministic {B}ell measurement using quantum
  communication bus},\ }\href@noop {} {\bibfield  {journal} {\bibinfo
  {journal} {Quantum Information Processing}\ }\textbf {\bibinfo {volume}
  {16}},\ \bibinfo {pages} {63} (\bibinfo {year} {2017})}\BibitemShut {NoStop}%
\bibitem [{\citenamefont {Zaidi}\ and\ \citenamefont {van
  Loock}(2013)}]{zaidi_prl_2013}%
  \BibitemOpen
  \bibfield  {author} {\bibinfo {author} {\bibfnamefont {H.~A.}\ \bibnamefont
  {Zaidi}}\ and\ \bibinfo {author} {\bibfnamefont {P.}~\bibnamefont {van
  Loock}},\ }\bibfield  {title} {\bibinfo {title} {Beating the one-half limit
  of ancilla-free linear optics {B}ell measurements},\ }\href
  {https://doi.org/10.1103/PhysRevLett.110.260501} {\bibfield  {journal}
  {\bibinfo  {journal} {Phys. Rev. Lett.}\ }\textbf {\bibinfo {volume} {110}},\
  \bibinfo {pages} {260501} (\bibinfo {year} {2013})}\BibitemShut {NoStop}%
\bibitem [{\citenamefont {Ewert}\ and\ \citenamefont {van
  Loock}(2014)}]{Ewert_prl_2014}%
  \BibitemOpen
  \bibfield  {author} {\bibinfo {author} {\bibfnamefont {F.}~\bibnamefont
  {Ewert}}\ and\ \bibinfo {author} {\bibfnamefont {P.}~\bibnamefont {van
  Loock}},\ }\bibfield  {title} {\bibinfo {title} {$3/4$-efficient {B}ell
  measurement with passive linear optics and unentangled ancillae},\ }\href
  {https://doi.org/10.1103/PhysRevLett.113.140403} {\bibfield  {journal}
  {\bibinfo  {journal} {Phys. Rev. Lett.}\ }\textbf {\bibinfo {volume} {113}},\
  \bibinfo {pages} {140403} (\bibinfo {year} {2014})}\BibitemShut {NoStop}%
\bibitem [{\citenamefont {Bayerbach}\ \emph {et~al.}(2023)\citenamefont
  {Bayerbach}, \citenamefont {D’Aurelio}, \citenamefont {van Loock},\ and\
  \citenamefont {Barz}}]{Matthias_SA_2023}%
  \BibitemOpen
  \bibfield  {author} {\bibinfo {author} {\bibfnamefont {M.~J.}\ \bibnamefont
  {Bayerbach}}, \bibinfo {author} {\bibfnamefont {S.~E.}\ \bibnamefont
  {D’Aurelio}}, \bibinfo {author} {\bibfnamefont {P.}~\bibnamefont {van
  Loock}},\ and\ \bibinfo {author} {\bibfnamefont {S.}~\bibnamefont {Barz}},\
  }\bibfield  {title} {\bibinfo {title} {Bell-state measurement exceeding 50\%
  success probability with linear optics},\ }\href
  {https://doi.org/10.1126/sciadv.adf4080} {\bibfield  {journal} {\bibinfo
  {journal} {Science Advances}\ }\textbf {\bibinfo {volume} {9}},\ \bibinfo
  {pages} {eadf4080} (\bibinfo {year} {2023})},\ \Eprint
  {https://arxiv.org/abs/https://www.science.org/doi/pdf/10.1126/sciadv.adf4080}
  {https://www.science.org/doi/pdf/10.1126/sciadv.adf4080} \BibitemShut
  {NoStop}%
\bibitem [{\citenamefont {van Loock}\ \emph {et~al.}(2006)\citenamefont {van
  Loock}, \citenamefont {Ladd}, \citenamefont {Sanaka}, \citenamefont
  {Yamaguchi}, \citenamefont {Nemoto}, \citenamefont {Munro},\ and\
  \citenamefont {Yamamoto}}]{vanLoock_PRL_2006}%
  \BibitemOpen
  \bibfield  {author} {\bibinfo {author} {\bibfnamefont {P.}~\bibnamefont {van
  Loock}}, \bibinfo {author} {\bibfnamefont {T.~D.}\ \bibnamefont {Ladd}},
  \bibinfo {author} {\bibfnamefont {K.}~\bibnamefont {Sanaka}}, \bibinfo
  {author} {\bibfnamefont {F.}~\bibnamefont {Yamaguchi}}, \bibinfo {author}
  {\bibfnamefont {K.}~\bibnamefont {Nemoto}}, \bibinfo {author} {\bibfnamefont
  {W.~J.}\ \bibnamefont {Munro}},\ and\ \bibinfo {author} {\bibfnamefont
  {Y.}~\bibnamefont {Yamamoto}},\ }\bibfield  {title} {\bibinfo {title} {Hybrid
  quantum repeater using bright coherent light},\ }\href
  {https://doi.org/10.1103/PhysRevLett.96.240501} {\bibfield  {journal}
  {\bibinfo  {journal} {Phys. Rev. Lett.}\ }\textbf {\bibinfo {volume} {96}},\
  \bibinfo {pages} {240501} (\bibinfo {year} {2006})}\BibitemShut {NoStop}%
\bibitem [{\citenamefont {Munro}\ \emph {et~al.}(2008)\citenamefont {Munro},
  \citenamefont {Van~Meter}, \citenamefont {Louis},\ and\ \citenamefont
  {Nemoto}}]{Munro_prl_2008}%
  \BibitemOpen
  \bibfield  {author} {\bibinfo {author} {\bibfnamefont {W.~J.}\ \bibnamefont
  {Munro}}, \bibinfo {author} {\bibfnamefont {R.}~\bibnamefont {Van~Meter}},
  \bibinfo {author} {\bibfnamefont {S.~G.~R.}\ \bibnamefont {Louis}},\ and\
  \bibinfo {author} {\bibfnamefont {K.}~\bibnamefont {Nemoto}},\ }\bibfield
  {title} {\bibinfo {title} {High-bandwidth hybrid quantum repeater},\ }\href
  {https://doi.org/10.1103/PhysRevLett.101.040502} {\bibfield  {journal}
  {\bibinfo  {journal} {Phys. Rev. Lett.}\ }\textbf {\bibinfo {volume} {101}},\
  \bibinfo {pages} {040502} (\bibinfo {year} {2008})}\BibitemShut {NoStop}%
\bibitem [{\citenamefont {Macridin}\ \emph {et~al.}(2022)\citenamefont
  {Macridin}, \citenamefont {Li}, \citenamefont {Mrenna},\ and\ \citenamefont
  {Spentzouris}}]{macridin_pra_2021}%
  \BibitemOpen
  \bibfield  {author} {\bibinfo {author} {\bibfnamefont {A.}~\bibnamefont
  {Macridin}}, \bibinfo {author} {\bibfnamefont {A.~C.~Y.}\ \bibnamefont {Li}},
  \bibinfo {author} {\bibfnamefont {S.}~\bibnamefont {Mrenna}},\ and\ \bibinfo
  {author} {\bibfnamefont {P.}~\bibnamefont {Spentzouris}},\ }\bibfield
  {title} {\bibinfo {title} {Bosonic field digitization for quantum
  computers},\ }\href {https://doi.org/10.1103/PhysRevA.105.052405} {\bibfield
  {journal} {\bibinfo  {journal} {Phys. Rev. A}\ }\textbf {\bibinfo {volume}
  {105}},\ \bibinfo {pages} {052405} (\bibinfo {year} {2022})}\BibitemShut
  {NoStop}%
\bibitem [{\citenamefont {Macridin}\ \emph {et~al.}(2024)\citenamefont
  {Macridin}, \citenamefont {Li},\ and\ \citenamefont
  {Spentzouris}}]{macridin_pra_2024}%
  \BibitemOpen
  \bibfield  {author} {\bibinfo {author} {\bibfnamefont {A.}~\bibnamefont
  {Macridin}}, \bibinfo {author} {\bibfnamefont {A.~C.~Y.}\ \bibnamefont
  {Li}},\ and\ \bibinfo {author} {\bibfnamefont {P.}~\bibnamefont
  {Spentzouris}},\ }\bibfield  {title} {\bibinfo {title} {Qumode transfer
  between continuous- and discrete-variable devices},\ }\href
  {https://doi.org/10.1103/PhysRevA.109.032419} {\bibfield  {journal} {\bibinfo
   {journal} {Phys. Rev. A}\ }\textbf {\bibinfo {volume} {109}},\ \bibinfo
  {pages} {032419} (\bibinfo {year} {2024})}\BibitemShut {NoStop}%
\bibitem [{\citenamefont {Bishop}\ \emph {et~al.}(2009)\citenamefont {Bishop},
  \citenamefont {Chow}, \citenamefont {Koch}, \citenamefont {Houck},
  \citenamefont {Devoret}, \citenamefont {Thuneberg}, \citenamefont {Girvin},\
  and\ \citenamefont {Schoelkopf}}]{Bishop2009}%
  \BibitemOpen
  \bibfield  {author} {\bibinfo {author} {\bibfnamefont {L.~S.}\ \bibnamefont
  {Bishop}}, \bibinfo {author} {\bibfnamefont {J.~M.}\ \bibnamefont {Chow}},
  \bibinfo {author} {\bibfnamefont {J.}~\bibnamefont {Koch}}, \bibinfo {author}
  {\bibfnamefont {A.~A.}\ \bibnamefont {Houck}}, \bibinfo {author}
  {\bibfnamefont {M.~H.}\ \bibnamefont {Devoret}}, \bibinfo {author}
  {\bibfnamefont {E.}~\bibnamefont {Thuneberg}}, \bibinfo {author}
  {\bibfnamefont {S.~M.}\ \bibnamefont {Girvin}},\ and\ \bibinfo {author}
  {\bibfnamefont {R.~J.}\ \bibnamefont {Schoelkopf}},\ }\bibfield  {title}
  {\bibinfo {title} {Nonlinear response of the vacuum {Rabi} resonance},\
  }\href {https://doi.org/10.1038/nphys1154} {\bibfield  {journal} {\bibinfo
  {journal} {Nature Phys.}\ }\textbf {\bibinfo {volume} {5}},\ \bibinfo {pages}
  {105} (\bibinfo {year} {2009})}\BibitemShut {NoStop}%
\bibitem [{\citenamefont {Walther}\ \emph {et~al.}(2006)\citenamefont
  {Walther}, \citenamefont {Varcoe}, \citenamefont {Englert},\ and\
  \citenamefont {Becker}}]{Walther2006}%
  \BibitemOpen
  \bibfield  {author} {\bibinfo {author} {\bibfnamefont {H.}~\bibnamefont
  {Walther}}, \bibinfo {author} {\bibfnamefont {B.~T.~H.}\ \bibnamefont
  {Varcoe}}, \bibinfo {author} {\bibfnamefont {B.-G.}\ \bibnamefont
  {Englert}},\ and\ \bibinfo {author} {\bibfnamefont {T.}~\bibnamefont
  {Becker}},\ }\bibfield  {title} {\bibinfo {title} {Cavity quantum
  electrodynamics},\ }\href {https://doi.org/10.1088/0034-4885/69/5/R02}
  {\bibfield  {journal} {\bibinfo  {journal} {Reports on Progress in Physics}\
  }\textbf {\bibinfo {volume} {69}},\ \bibinfo {pages} {1325} (\bibinfo {year}
  {2006})}\BibitemShut {NoStop}%
\bibitem [{\citenamefont {Cottet}\ \emph {et~al.}(2017)\citenamefont {Cottet},
  \citenamefont {Dartiailh}, \citenamefont {Desjardins}, \citenamefont
  {Cubaynes}, \citenamefont {Contamin}, \citenamefont {Delbecq}, \citenamefont
  {Viennot}, \citenamefont {Bruhat}, \citenamefont {Douçot},\ and\
  \citenamefont {Kontos}}]{Cottet2017}%
  \BibitemOpen
  \bibfield  {author} {\bibinfo {author} {\bibfnamefont {A.}~\bibnamefont
  {Cottet}}, \bibinfo {author} {\bibfnamefont {M.~C.}\ \bibnamefont
  {Dartiailh}}, \bibinfo {author} {\bibfnamefont {M.~M.}\ \bibnamefont
  {Desjardins}}, \bibinfo {author} {\bibfnamefont {T.}~\bibnamefont
  {Cubaynes}}, \bibinfo {author} {\bibfnamefont {L.~C.}\ \bibnamefont
  {Contamin}}, \bibinfo {author} {\bibfnamefont {M.}~\bibnamefont {Delbecq}},
  \bibinfo {author} {\bibfnamefont {J.~J.}\ \bibnamefont {Viennot}}, \bibinfo
  {author} {\bibfnamefont {L.~E.}\ \bibnamefont {Bruhat}}, \bibinfo {author}
  {\bibfnamefont {B.}~\bibnamefont {Douçot}},\ and\ \bibinfo {author}
  {\bibfnamefont {T.}~\bibnamefont {Kontos}},\ }\bibfield  {title} {\bibinfo
  {title} {Cavity qed with hybrid nanocircuits: from atomic-like physics to
  condensed matter phenomena},\ }\href
  {https://doi.org/10.1088/1361-648X/aa7b4d} {\bibfield  {journal} {\bibinfo
  {journal} {Journal of Physics: Condensed Matter}\ }\textbf {\bibinfo {volume}
  {29}},\ \bibinfo {pages} {433002} (\bibinfo {year} {2017})}\BibitemShut
  {NoStop}%
\bibitem [{\citenamefont {Blais}\ \emph {et~al.}(2021)\citenamefont {Blais},
  \citenamefont {Grimsmo}, \citenamefont {Girvin},\ and\ \citenamefont
  {Wallraff}}]{Blais2021}%
  \BibitemOpen
  \bibfield  {author} {\bibinfo {author} {\bibfnamefont {A.}~\bibnamefont
  {Blais}}, \bibinfo {author} {\bibfnamefont {A.~L.}\ \bibnamefont {Grimsmo}},
  \bibinfo {author} {\bibfnamefont {S.~M.}\ \bibnamefont {Girvin}},\ and\
  \bibinfo {author} {\bibfnamefont {A.}~\bibnamefont {Wallraff}},\ }\bibfield
  {title} {\bibinfo {title} {Circuit quantum electrodynamics},\ }\href
  {https://doi.org/10.1103/RevModPhys.93.025005} {\bibfield  {journal}
  {\bibinfo  {journal} {Rev. Mod. Phys.}\ }\textbf {\bibinfo {volume} {93}},\
  \bibinfo {pages} {025005} (\bibinfo {year} {2021})}\BibitemShut {NoStop}%
\bibitem [{\citenamefont {Gerry}\ and\ \citenamefont
  {Knight}(2004)}]{gerry_knight_2004}%
  \BibitemOpen
  \bibfield  {author} {\bibinfo {author} {\bibfnamefont {C.}~\bibnamefont
  {Gerry}}\ and\ \bibinfo {author} {\bibfnamefont {P.}~\bibnamefont {Knight}},\
  }\href {https://doi.org/10.1017/CBO9780511791239} {\emph {\bibinfo {title}
  {Introductory Quantum Optics}}}\ (\bibinfo  {publisher} {Cambridge University
  Press, New York},\ \bibinfo {year} {2004})\BibitemShut {NoStop}%
\bibitem [{\citenamefont {Vahlbruch}\ \emph {et~al.}(2016)\citenamefont
  {Vahlbruch}, \citenamefont {Mehmet}, \citenamefont {Danzmann},\ and\
  \citenamefont {Schnabel}}]{Vahlbruch_prl_2016}%
  \BibitemOpen
  \bibfield  {author} {\bibinfo {author} {\bibfnamefont {H.}~\bibnamefont
  {Vahlbruch}}, \bibinfo {author} {\bibfnamefont {M.}~\bibnamefont {Mehmet}},
  \bibinfo {author} {\bibfnamefont {K.}~\bibnamefont {Danzmann}},\ and\
  \bibinfo {author} {\bibfnamefont {R.}~\bibnamefont {Schnabel}},\ }\bibfield
  {title} {\bibinfo {title} {Detection of 15 db squeezed states of light and
  their application for the absolute calibration of photoelectric quantum
  efficiency},\ }\href {https://doi.org/10.1103/PhysRevLett.117.110801}
  {\bibfield  {journal} {\bibinfo  {journal} {Phys. Rev. Lett.}\ }\textbf
  {\bibinfo {volume} {117}},\ \bibinfo {pages} {110801} (\bibinfo {year}
  {2016})}\BibitemShut {NoStop}%
\bibitem [{\citenamefont {Nielsen}\ and\ \citenamefont
  {Chuang}(2010)}]{nielsen2010quantum}%
  \BibitemOpen
  \bibfield  {author} {\bibinfo {author} {\bibfnamefont {M.~A.}\ \bibnamefont
  {Nielsen}}\ and\ \bibinfo {author} {\bibfnamefont {I.}~\bibnamefont
  {Chuang}},\ }\href@noop {} {\bibinfo {title} {Quantum computation and quantum
  information: 10th anniversary edition}} (\bibinfo {year} {2010})\BibitemShut
  {NoStop}%
\bibitem [{\citenamefont {Sch\"{o}nbeck}\ \emph {et~al.}(2018)\citenamefont
  {Sch\"{o}nbeck}, \citenamefont {Thies},\ and\ \citenamefont
  {Schnabel}}]{Schonbeck_2018}%
  \BibitemOpen
  \bibfield  {author} {\bibinfo {author} {\bibfnamefont {A.}~\bibnamefont
  {Sch\"{o}nbeck}}, \bibinfo {author} {\bibfnamefont {F.}~\bibnamefont
  {Thies}},\ and\ \bibinfo {author} {\bibfnamefont {R.}~\bibnamefont
  {Schnabel}},\ }\href {https://doi.org/10.1364/OL.43.000110} {\bibfield
  {journal} {\bibinfo  {journal} {Opt. Lett.}\ }\textbf {\bibinfo {volume}
  {43}},\ \bibinfo {pages} {110} (\bibinfo {year} {2018})}\BibitemShut
  {NoStop}%
\bibitem [{\citenamefont {Hagemann}\ \emph {et~al.}(2024)\citenamefont
  {Hagemann}, \citenamefont {Zander}, \citenamefont {Sch\"{o}nbeck},\ and\
  \citenamefont {Schnabel}}]{Hagemann_2024}%
  \BibitemOpen
  \bibfield  {author} {\bibinfo {author} {\bibfnamefont {M.}~\bibnamefont
  {Hagemann}}, \bibinfo {author} {\bibfnamefont {J.}~\bibnamefont {Zander}},
  \bibinfo {author} {\bibfnamefont {A.}~\bibnamefont {Sch\"{o}nbeck}},\ and\
  \bibinfo {author} {\bibfnamefont {R.}~\bibnamefont {Schnabel}},\ }\bibfield
  {title} {\bibinfo {title} {10-db squeeze laser tuneable over half a nanometer
  around 1550 nm},\ }\href {https://doi.org/10.1364/OE.507573} {\bibfield
  {journal} {\bibinfo  {journal} {Opt. Express}\ }\textbf {\bibinfo {volume}
  {32}},\ \bibinfo {pages} {7954} (\bibinfo {year} {2024})}\BibitemShut
  {NoStop}%
\bibitem [{\citenamefont {Li}\ \emph {et~al.}(2022)\citenamefont {Li},
  \citenamefont {Wang}, \citenamefont {You},\ and\ \citenamefont
  {Zhu}}]{Li_2022}%
  \BibitemOpen
  \bibfield  {author} {\bibinfo {author} {\bibfnamefont {J.}~\bibnamefont
  {Li}}, \bibinfo {author} {\bibfnamefont {Y.-P.}\ \bibnamefont {Wang}},
  \bibinfo {author} {\bibfnamefont {J.-Q.}\ \bibnamefont {You}},\ and\ \bibinfo
  {author} {\bibfnamefont {S.-Y.}\ \bibnamefont {Zhu}},\ }\bibfield  {title}
  {\bibinfo {title} {{Squeezing microwaves by magnetostriction}},\ }\href@noop
  {} {\bibfield  {journal} {\bibinfo  {journal} {National Science Review}\
  }\textbf {\bibinfo {volume} {10}},\ \bibinfo {pages} {247} (\bibinfo {year}
  {2022})}\BibitemShut {NoStop}%
\end{thebibliography}

\appendix

\section{CV entanglement swapping}
\label{app:CVent}

The continuous-variable entanglement swapping protocol was introduced in~\cite{Look_PRA_1999}. Here, we present a modified version similar to our hybrid protocol. In fact, the hybrid protocol, in the limit of large squeezing and large qubit register sizes, converges to the continuous-variable protocol
presented here.

\begin{enumerate}
\item
Alice starts with two independent modes, $1$ and $2$:
\begin{align}
\label{eq:cvesp_A1}
\ket{\chi\left(\text{step1}\right)}_{A} = \int \ket{x_1}_A dx_1 \otimes \int \ket{x_2}_A dx_2
\end{align}
Similarly, Bob starts with the modes $3$ and $4$:
\begin{align}
\label{eq:cvesp_B1}
\ket{\chi\left(\text{step1}\right)}_{B} = \int \ket{x_3}_B dx_3 \otimes \int \ket{x_4}_B dx_4
\end{align}
The wavefunctions of the modes in~\cref{eq:cvesp_A1,eq:cvesp_B1} are unnormalized and describe infinitely squeezed vacuum states, as they can be obtained by taking $\sigma \rightarrow \infty$ in~\cref{eq:gfun}.

\item
Alice applies the operator $e^{-i X_1 \otimes X_2}$:
\begin{align}
\ket{\chi\left(\text{step2}\right)}_{A}=\int e^{-i x_1 x_2}\ket{x_1, x_2}_A dx_1 dx_2.
\end{align}
Similarly,  Bob  applies the operator $e^{-i X_3 \otimes X_4}$:
\begin{align}
\ket{\chi\left(\text{step2}\right)}_{B}=\int e^{-i x_3 x_4}\ket{x_3, x_4}_B dx_3 dx_4.
\end{align}
\item
Alice sends mode $2$ to Charlie, and Bob sends mode $3$ to Charlie. Charlie sends the modes $2$ and $3$  thorough a $50:50$ beam splitter,
\begin{align}
\ket{\chi\left(\text{step3}\right)}_{ABC}=\int dx_1 dx_2 dx_3 dx_4 e^{-i x_1 x_2} e^{-i x_3 x_4} \ket{x_1, x_4}_{AB} \ket{\frac{x_2+x_3}{\sqrt{2}}, \frac{-x_2+x_3}{\sqrt{2}}}_C.
\end{align}
\item
Employing homodyne measurement, Charlie measures the quadrature $X$ for one mode and the quadrature $P$ for the other mode. Denoting the measurement results by $x_D$ and $p_D$, the state becomes (see~\cref{eq:homodyne}):
\begin{align}
\ket{\chi\left(\text{step4}\right)}_{AB}&=\int dx_1 dx_4 \ket{x_1, x_4}_{AB} \frac{1}{\sqrt{2 \pi}}\int dx_2 dx_3 e^{-i \left(x_1 x_2+x_3 x_4\right)} \delta\left(x_D-\frac{x_2+x_3}{\sqrt{2}}\right)e^{-i p_D \frac{-x_2+x_3}{\sqrt{2}}} \\ \nonumber
% &=\int dx_1 dx_4 \ket{x_1, x_4} \frac{1}{\sqrt{\pi}}\int dx_2 dx_3 e^{-i \left(x_1 x_2+x_3 x_4\right)} \delta\left(\sqrt{2} x_D-x_2-x_3\right)e^{-i p_D \frac{-x_2+x_3}{\sqrt{2}}}\\ \nonumber
% &=e^{-i p_D x_D}\int dx_1 dx_4 \ket{x_1, x_4} e^{-i \sqrt{2} x_D x_4}\frac{1}{\sqrt{\pi}}\int dx_2 e^{i \sqrt{2} p_D x_2} e^{-i x_1 x_2} e^{i x_2 x_4}
% \\ \nonumber
% &=e^{-i p_D x_D}\int dx_1 dx_4 \ket{x_1, x_4} e^{-i \sqrt{2} x_D x_4}\frac{1}{\sqrt{\pi}}\int dx_2 e^{i \left(\sqrt{2} p_D-x_1+x_4 \right) x_2 }
% \\ \nonumber
% &=e^{-i p_D x_D}\int dx_1 dx_4 \ket{x_1, x_4} e^{-i \sqrt{2} x_D x_4} \sqrt{2} \delta \left(\sqrt{2} p_D-x_1+x_4\right)\\ \nonumber
&=\sqrt{2} e^{-i p_D x_D} \int dx \ket{x+\sqrt{2} p_D, x }_{AB}e^{-i \sqrt{2} x_D x}
\end{align}

\item
Bob applies the displacement operator $e^{i \sqrt{2} x_D X_4}$:
\begin{align}
\ket{\chi\left(\text{step5}\right)}_{AB}&=\sqrt{2} e^{-i p_D x_D} \int dx \ket{x+\sqrt{2} p_D, x }_{AB}.
\end{align}
\item
Alice applies the displacement operator $e^{i \sqrt{2} p_D P_1}$:
\begin{align}
\ket{\chi\left(\text{step6}\right)}_{AB}&=\sqrt{2} e^{-i p_D x_D} \int dx \ket{x, x }_{AB}.
\end{align}
\end{enumerate}
To conclude, by performing homodyne measurements at Charlie, the protocol produces an entangled state between the  mode $1$ at Alice and mode $4$ at Bob.

\section{Homodyne measurement}
\label{app:homodyne}

The result of the homodyne measurement is obtained by applying the projector
operator $\ketbra{x_D,p_D}{x_D,p_D}$ to the state described by~\cref{eq:st3}.
Considering that
\begin{align}
\label{eq:homodyne}
\braket{x_D,p_D}{\frac{x_2+x_3}{\sqrt{2}}, \frac{-x_2+x_3}{\sqrt{2}}}=\delta\left(x_D-\frac{x_2+x_3}{\sqrt{2}}\right) \frac{1}{\sqrt{2 \pi}} e^{-i p_D \frac{-x_2+x_3}{\sqrt{2}}},
\end{align}
\noindent the post-measurement state becomes:
\begin{align}
\ket{\chi\left(\text{step4}\right)}_{AB}
%&= \frac{1}{2^n \sqrt{P\left(x_D,p_D\right)}} \sum_{j,k=0}^{2^n-1} \ket{j, k}_{AB} \frac{1}{\sqrt{2 \pi}}\int dx_2 dx_3 g(x_2) g(x_3) e^{-i \left(\x_j x_2+x_3 \x_k\right)} \delta\left(x_D-\frac{x_2+x_3}{\sqrt{2}}\right)e^{-i p_D \frac{-x_2+x_3}{\sqrt{2}}} \\ \nonumber
&=\frac{1}{2^n \sqrt{P\left(x_D,p_D\right)}} \sum_{j,k=0}^{2^n-1} \ket{j, k}_{AB} \frac{1}{\sqrt{\pi}}\int dx_2 dx_3 e^{-i \left(\x_j x_2+x_3 \x_k\right)} g(x_2) g(x_3) \delta\left(\sqrt{2} x_D-x_2-x_3\right)e^{-i p_D \frac{-x_2+x_3}{\sqrt{2}}}\\ \nonumber
% &=e^{-i p_D x_D}\frac{1}{2^n \sqrt{P\left(x_D,p_D\right)}} \sum_{j,k=0}^{2^n-1} \ket{j, k}_{AB} e^{-i \sqrt{2} x_D \x_k}\frac{1}{\sqrt{\pi}}\int dx_2 g(x_2) g(\sqrt{2} x_D-x_2) e^{i \sqrt{2} p_D x_2} e^{-i \x_j x_2} e^{i x_2 \x_k}
% \\ \nonumber
&=e^{-i p_D x_D}\frac{1}{2^n \sqrt{P\left(x_D,p_D\right)}} \sum_{j,k=0}^{2^n-1} \ket{j, k}_{AB} e^{-i \sqrt{2} x_D \x_k}\frac{1}{\sqrt{\pi}}\int dx_2 g(x_2) g(\sqrt{2} x_D-x_2) e^{i \left(\sqrt{2} p_D-\x_j+\x_k \right) x_2 }
\\ \nonumber
% &=e^{-i p_D x_D}\frac{1}{2^n \sqrt{P\left(x_D,p_D\right)}} \sum_{j,k=0}^{2^n-1} \ket{j, k}_{AB} e^{-i \sqrt{2} x_D \x_k}\frac{1}{\sqrt{2\pi}} e^{i \left(\sqrt{2} p_D-\x_j+\x_k \right) \frac{x_D}{\sqrt{2}} }\int dy g(\frac{x_D+y}{\sqrt{2}}) g(\frac{x_D-y}{\sqrt{2}}) e^{i \left(p_D-\frac{\x_j-\x_k}{\sqrt{2}} \right) y} \\ \nonumber
&=\frac{1}{2^n \sqrt{P\left(x_D,p_D\right)}} \sum_{j,k=0}^{2^n-1} e^{-i x_D \frac{\x_j+\x_k }{\sqrt{2}} } \ket{j, k}_{AB} \frac{1}{\sqrt{2\pi}}\int dy g(\frac{x_D+y}{\sqrt{2}}) g(\frac{x_D-y}{\sqrt{2}}) e^{i \left(p_D-\frac{\x_j-\x_k}{\sqrt{2}} \right) y}\\ \nonumber
&= \frac{1}{2^n \sqrt{P\left(x_D,p_D\right)}}
\sum_{j,k=0}^{2^n-1}
e^{-i x_D \frac{\x_j+\x_k }{\sqrt{2}}}  \hh_{x_D}\left(p_D-\frac{\x_j-\x_k}{\sqrt{2}}\right)\ket{j, k}_{AB},
\end{align}
\noindent where $\hh_{x_D}(p)$ is defined by ~\cref{eq:hhfun}.

\end{document}